\documentclass[12pt]{article}

\usepackage{fullpage}
\usepackage{cite}
\usepackage{amsmath}
\usepackage{amssymb}

\usepackage{epsfig}
\usepackage{psfrag}
\usepackage{graphicx}
\usepackage[utf8]{inputenc}

\title{}
\date{}
\def\para{\\ [-2mm]}

\def\para{\\ [-2mm]}
\def \be  {\begin{equation}}
\def \ee  {\end{equation}}
\def \ba  {\begin{eqnarray}}
\def \ea  {\end{eqnarray}}
\newcommand{\nn}{\nonumber}
\def\eqn#1{eq.~(\ref{#1})} 

\def\eqns#1#2{eqs.~(\ref{#1}) and~(\ref{#2})}

\def\IZ{\relax\ifmmode\mathchoice
{\hbox{\cmss Z\kern-.4em Z}}{\hbox{\cmss Z\kern-.4em Z}}
{\lower.4pt\hbox{\cmsss Z\kern-.4em Z}}
{\lower1.2pt\hbox{\cmsss Z\kern-.4em Z}}\else{\cmss Z\kern-.4em Z}\fi}
\newcommand{\Z}{\mathsf{Z}\kern -5pt \mathsf{Z}}
\newcommand{\unit}{\mathsf{1}\kern -3pt \mathsf{l}}

\def \Tr {\mathop{\rm Tr}\nolimits}

\def\fr#1#2{ {\textstyle{#1 \over #2}}}

\def\de {\epsilon}

\def\eps{\epsilon}

\def\cA {  {\cal A} }

\def\cM {  {\cal M} }
\def\cN {  {\cal N} }
\def\cO {  {\cal O} }
\def\< { \langle}
\def\> { \rangle}
\def\Zero{ { (0) }}
\def\One {{ (1) }}
\def\Two{ {(2)} }
\def\Three{{(3)} }

\def\Ell{{(\ell)}}

\renewcommand{\vec}[1]{{\bf #1} }

\def\regge{ \quad \longrightarrow \quad}

\def\tf{\tilde{f}}
\def\lam{\lambda}
\def\oneN{ {1 \over N} }
\def\Ell{{(\ell)}}
\def\Ellk{{(\ell,k)}}
\def\Elltwok{{(\ell,2k)}}
\def\Elltwokplus{{(\ell,2k+1)}}

\def\x{ {(x)} }
\def\xzero{ {(x,0)} }
\def\xone{ {(x,1)} }
\def\xtwo{ {(x,2)} }
\def\lad{ { ({\rm lad})  } }
\def\bx{ { ({\rm box})  } }
\def\P{{(P)}}
\def\Pzero{{(P,0)}}
\def\Pone{{(P,1)}}
\def\Ptwo{{(P,2)}}
\def\NP{{(NP)}}
\def\NPzero{{(NP,0)}}
\def\NPone{{(NP,1)}}
\def\NPtwo{{(NP,2)}}

\begin{document}

\titlepage
\begin{flushright}
BOW-PH-170\\
\end{flushright}

\vspace{3mm}

\begin{center}

{\Large\bf\sf
Proof of a three-loop relation between
\\ [2mm]
the Regge limits of four-point amplitudes 
\\ [2mm]
in $\cN=4$ SYM  
and $\cN=8$ supergravity 
\\ [4mm]
}

\vskip 3cm

{\sc
Stephen G. Naculich
and 
Theodore W. Wecker
}

\vskip 0.5cm
{\it
Department of Physics and Astronomy\\
Bowdoin College\\
Brunswick, ME 04011 USA
}

\vspace{5mm}
{\tt
naculich@bowdoin.edu, twecker@bowdoin.edu
}
\end{center}

\vskip 3cm

\begin{abstract}

A previously proposed 
all-loop-orders relation
between the Regge limits of 
four-point amplitudes of
$\cN=4$ supersymmetric Yang-Mills theory 
and  $\cN=8$ supergravity 
is established at the three-loop level.
We show that the Regge limit of known expressions 
for the amplitudes obtained using generalized unitarity
simplifies in both cases to a (modified) sum over 
three-loop ladder and crossed-ladder scalar diagrams.
This in turn is consistent with the result obtained 
using the eikonal representation of the four-point gravity amplitude.
A possible exact three-loop relation 
between four-point amplitudes is also considered.
\end{abstract}

\vspace*{0.5cm}

\vfil\break

\section{Introduction}
\setcounter{equation}{0}

In recent years, 
connections between the perturbative amplitudes
of gauge theory and gravity have been 
explored intensively
(see ref.~\cite{Bern:2019prr} for a recent review).
The first hints of such a connection came from string theory,
when Kawai, Lewellen, and Tye obtained a relation 
between open and closed supersymmetric string amplitudes 
at tree level \cite{Kawai:1985xq}.
In the low energy limit, 
this implies relations between tree-level gauge-theory and gravity amplitudes.
\para

The most comprehensive current understanding of such relations 
is through the double copy of Bern, Carrasco, and 
Johansson \cite{Bern:2008qj,Bern:2010ue}.
When gauge-theory amplitudes are written in terms of 
color factors and kinematic numerators of graphs with trivalent vertices,
with the kinematic numerators 
obeying the same algebraic relations 
as those satisfied by the color factors (color-kinematic duality),
then gravity amplitudes can be obtained 
from the gauge-theory amplitudes
by replacing the color factors 
with a second copy of the kinematic numerators.
That this procedure gives correct gravity amplitudes 
was proven at tree level in ref.~\cite{Bern:2010yg}.
At higher loops, the gauge-theory amplitudes are written in 
terms of integrals over loop momenta of graphs with trivalent vertices,
with the kinematic numerators of the integrands
chosen to obey color-kinematic duality.
That the double-copy procedure 
generates correct loop-level gravity amplitudes
was demonstrated through four loops 
for four-point amplitudes of $\cN=8$ supergravity
\cite{Bern:2010ue,Bern:2012uf}.
At five loops and above, it is difficult to find representations 
of the $\cN=4$ SYM four-point amplitude with manifest color-kinematic duality, 
but double-copy representations of 
the $\cN=8$ supergravity four-point amplitude 
have nonetheless been obtained \cite{Bern:2017ucb,Bern:2018jmv}.
\para

It is important to emphasize that, at loop level, 
the double-copy prescription applies at the level of {\it integrands}.
The double copy does not imply direct relations 
between the {\it integrated amplitudes} of gauge theory and gravity 
except in those cases (e.g., one- and two-loop four-point amplitudes,
or one-loop five-point amplitudes)
in which kinematic numerators are independent 
of loop momenta
and therefore can be pulled outside the 
integrals \cite{
Carrasco:2011mn,
Bern:2011rj,
Naculich:2011fw,
BoucherVeronneau:2011qv,
Naculich:2011my}.
\para

A higher-loop relation between gauge-theory 
and gravity amplitudes was recently conjectured to hold
in the Regge limit by one of the authors \cite{Naculich:2020clm}. 
In that paper, 
the Regge limit of (nonplanar) 
$\ell$-loop $\cN=4$ SYM four-point amplitudes 
was examined,
and a basis of color factors suitable for that limit
was presented.
The coefficients of the four-point amplitude 
in that basis were calculated through three-loop order,
using the Regge limit of the full amplitude
previously obtained by Henn and Mistlberger \cite{Henn:2016jdu}.
One of those coefficients, 
denoted $B_{\ell\ell}^\Ell$, 
whose Laurent expansion begins at $1/\de^\ell$, 
was shown to be proportional to the Regge limit of the 
$\ell$-loop  $\cN=8$ supergravity four-point amplitude,
at least through the first three orders in the Laurent expansion,
thus motivating the conjecture.
\para

At one and two loops, 
the conjectured relation reads
\begin{align}
{1 \over   (\kappa_D/2)^2  s t u} {\cM^\One \over \cM^\Zero  }
&\regge
~-~ { \left[ A_1^{(1)}  + A_3^{(1)}\right]
\over g_D^2 s t A_1^\Zero }\,,
\label{onelooprelation}
\\
 {1 \over ( \kappa_D/2)^4    ~s t u }
 {\cM^\Two \over \cM^\Zero }
&\regge
~- {s \over 6} 
~ { \left[ A_7^\Two  - A_9^\Two  \right] \over g_D^4  s t A_1^\Zero  }\,.
\label{twolooprelation}
\end{align}

These are in fact a specialization 
to the Regge limit of previously known
\cite{Green:1982sw,Naculich:2008ys}
exact relations (i.e. relations that hold in all kinematic regions),
namely \eqns{exactonelooprelation}{exacttwolooprelation},
as will be shown explicitly  below.
Here $\cM^\Ell$ denotes the $\ell$-loop four-point amplitude 
of $\cN=8$ supergravity, while 
$A^\Ell_\lam$ denote the color-ordered $\ell$-loop four-point amplitudes 
of $\cN=4$ SYM theory 
(i.e., the coefficients of the $\ell$-loop amplitude 
in the $(3\ell+3)$-dimensional extended trace basis $t_\lam^\Ell$
defined in sec.~\ref{sec:general}).
For each theory,
the loop- and tree-level amplitudes 
carry the same helicity dependence,
so the ratios $\cM^\Ell/\cM^\Zero$ and $A_\lam^\Ell/A_1^\Zero$ are 
helicity-independent functions of the Mandelstam variables $s$, $t$, and $u$.
Since loop-level amplitudes of massless particles are IR-divergent, 
we dimensionally regularize them in $D=4-2\de$ dimensions,
with $g_D$ and $\kappa_D$ denoting the 
gauge and gravitational couplings 
in $D$ dimensions,\footnote{Our convention is $(\kappa_D/2)^2 = 8 \pi G_D$.}
and the amplitudes are expressed as Laurent expansions in $\de$.
Here and throughout this paper, the long right arrow 
denotes a relation valid in the Regge limit $|t| \ll  s$. 
\para

The conjectured relation at three loops
\begin{align}
 {1 \over ( \kappa_D/2)^6    ~s t u }
 {\cM^\Three \over \cM^\Zero }
&\regge
~{s^2 \over 12} 
{\left[
  4 \left(  A_1^\Three + A_3^\Three \right) 
- \left( A_4^\Three +A_6^\Three 
       + A_7^\Three +A_9^\Three \right)
\right]
 \over
g_D^6  s t A_1^\Zero  }
\label{threelooprelation}
\end{align}

was shown \cite{Naculich:2020clm}
to hold through $\cO(\de^0)$ of the Laurent expansion
(i.e., the first four non-vanishing terms)
using the explicit calculations of 
Henn and Mistlberger \cite{Henn:2016jdu,Henn:2019rgj}.
The two- and three-loop relations,
\eqns{twolooprelation}{threelooprelation},
are special cases of the more general all-loop 
conjecture, namely
\begin{align}
 {1 \over ( \kappa_D/2)^{2\ell}     ~s t u }
 {\cM^\Ell \over \cM^\Zero }
&\regge
 ~-\frac{s^{\ell-1}}{2 \cdot 3^{\ell/2} }
~{ \left[ A_{3\ell+1}^\Ell - A_{3 \ell+3}^\Ell \right] 
\over 
g_D^{2\ell} s t  A_1^\Zero }  ,
\hskip25mm  \hbox{even } \ell \ge 2\,,
\label{evenloop}
\\
 {1 \over ( \kappa_D/2)^{2\ell}     ~s t u }
 {\cM^\Ell \over \cM^\Zero }
&\regge
\frac{ s^{\ell-1}}{ 4 \cdot  3^{(\ell-1)/2}   }
~ 
\frac{
\left[ 4 \left( A_{3\ell-8}^\Ell + A_{3 \ell-6}^\Ell \right)
- \left( A_{3\ell-5}^\Ell + A_{3 \ell-3}^\Ell +    A_{3\ell-2}^\Ell + A_{3 \ell}^\Ell\right)
\right]	
}
{g_D^{2\ell} s t  A_1^\Zero }  ,
\nn\\
&
\hskip95mm \hbox{odd } \ell \ge 3 \,.
\label{oddloop}
\end{align}

In ref.~\cite{Naculich:2020clm},
these relations were verified to hold at all loop orders
for the first three (IR-divergent) terms in the Laurent expansion, 
using the known structure of IR divergences in both theories.
\vfil\break

In ref.~\cite{Naculich:2020clm}, 
it was suggested that the three-loop relation (\ref{threelooprelation})
could be established to all orders in the Laurent expansion 
by examining known expressions for the amplitudes 
in terms of scalar integrals obtained through generalized unitarity.
That is the main aim of this paper.
We will demonstrate that both sides of \eqn{threelooprelation}
simplify in the Regge limit 
to the same (modified) sum over 
three-loop ladder and crossed-ladder scalar diagrams,
thus proving the conjectured relation.
The sum is modified in the sense that two of the 
crossed-ladder diagrams are multiplied by a factor of one-half
relative to the remaining diagrams.
\para

Alternatively, the eikonal approximation 
\cite{Cheng:1969eh,Abarbanel:1969ek,Levy:1969cr}
may be used to obtain
a representation (\ref{Meik}) of the supergravity amplitude
as an integral over impact-parameter space
\cite{tHooft:1987vrq,Amati:1987wq,Muzinich:1987in,Amati:1987uf,Kabat:1992tb,
Giddings:2010pp,Melville:2013qca,Akhoury:2013yua,Luna:2015paa,
DiVecchia:2019myk,DiVecchia:2019kta}.
This may then be evaluated to give an explicit result (\ref{Mell}) 
for the $\ell$-loop supergravity amplitude
in the Regge limit \cite{DiVecchia:2019myk}.
We show that this result is consistent with the
representation obtained in the current paper
of the Regge limit of the $\cN=8$ supergravity amplitude
at one, two, and three loops as a sum of 
ladder and crossed-ladder scalar diagrams.
The modification of the sum at three loops mentioned above 
is crucial for this consistency.
\para

In this paper, we also observe that 
the three-loop relation (\ref{threelooprelation})
is the Regge limit of a certain exact relation 
(\ref{exactthreelooprelation})
that would be valid if only a certain subset of the scalar diagrams
were included in the evaluation of the three-loop amplitudes.   
Testing this exact relation against the Laurent expansions
of the full three-loop amplitudes,
we find that it holds at $\cO(1/\de^3)$ and $\cO(1/\de^2)$,
only breaking down at $\cO(1/\de)$, cf. \eqn{breakdown}.
\para

The outline of this paper is as follows:
in section \ref{sec:general}
we review the form of maximally supersymmetric four-point amplitudes
obtained from generalized unitarity, 
and the definition of color-ordered amplitudes in the extended trace basis.
In sections \ref{sec:oneloop} and \ref{sec:twoloop} we write down the
one- and two-loop amplitudes for $\cN=4$ SYM and $\cN=8$ supergravity
in terms of scalar integrals,
the exact relations that hold among them, 
and the Regge limits of these relations.
In section \ref{sec:threeloop} we write down the
three-loop $\cN=4$ SYM and $\cN=8$ supergravity amplitudes
in terms of scalar integrals,
and then obtain their approximate form in the Regge limit,
thus establishing the three-loop relation (\ref{threelooprelation}).
We also present a putative exact three-loop relation,
and show that it only breaks down at $\cO(1/\de)$.
In section \ref{sec:ladders}, we show that the expressions obtained
in the previous three sections can be recast as a (modified) sum 
over ladder and crossed-ladder scalar diagrams, 
and show how this is related to the eikonal representation
of gravity amplitudes in impact-parameter space.
Section \ref{sec:concl} summarizes the results of the paper.

\section{Maximally supersymmetric four-point amplitudes}
\setcounter{equation}{0}
\label{sec:general}

Generalized unitarity \cite{Bern:1994zx,Bern:1994cg}
has been used to find representations of 
various loop-level amplitudes of maximally supersymmetric 
$\cN=4$ SYM and $\cN=8$ supergravity theories
in terms of planar and nonplanar scalar 
integrals \cite{Bern:1997nh,Bern:1998ug,Bern:2007hh,Bern:2008pv,Bern:2010tq, 
Carrasco:2011mn,
Bern:2012uf,Bern:2012uc,Bern:2017ucb,Bern:2018jmv}.
In this section, we review the results of this approach,
and establish our conventions for integrals and color-ordered amplitudes.

\subsection{$\cN=4$ SYM four-point amplitude}

The $\ell$-loop $\cN=4$ SYM four-point amplitude
can be expressed as a linear combination 
of products of 
color factors $c^\x$ and scalar integrals  $I^\xone$ 
associated with a set of diagrams $x$ with trivalent vertices.
The color factor $c^\x$ associated with each diagram is 
defined by decorating each vertex of the diagram
with a structure constant $\tf^{abc}$, 
related to the conventionally defined structure constants $f^{abc}$ by
\be
\tilde{f}^{abc} \equiv  i \sqrt{2} f^{abc} 
\label{tildef}
\ee
and then contracting indices connected by internal lines. 
The scalar integral 
$I^\xone$ associated with the diagram $x$ is defined 
(following the conventions of eq.~(2.4) of ref.~\cite{Bern:2010tq})
as 
\be
I^\xone (p_1,p_2,p_3,p_4) 
= (-i)^\ell \int \prod_{i=1}^\ell {d^D \ell_i \over (2\pi)^D} 
{N^\xone  \over \prod_{j=1}^{3\ell+1} l_j^2 }
\label{Ixone}
\ee

where $N^\xone$ is a specified numerator factor for the diagram.
It will later be convenient for us to define 
the related scalar integral $I^\xzero$,  without a numerator factor,
\be
I^\xzero (p_1,p_2,p_3,p_4) 
= (-i)^\ell \int \prod_{i=1}^\ell {d^D \ell_i \over (2\pi)^D} 
{1\over \prod_{j=1}^{3\ell+1} l_j^2 } \,.
\label{Ixzero}
\ee

The $\ell$-loop amplitude is then given by a sum over all the diagrams 
and, to ensure Bose symmetry, a sum  over all permutations of the 
external legs 
\begin{align}
\cA^\Ell \quad \sim \quad \sum_{x} \sum_{S_4} \lambda^\x  c_{ijkl}^{(x)} I_{ijkl}^\xone
\label{sumoverperms}
\end{align}
where $S_4$ denotes that $ijkl$ 
runs over all permutations of $1234$,
the $\lambda^{(x)}$ are simple combinatorial factors,
and we define
\be
I_{ijkl}^\x \equiv I^\x (p_i,p_j,p_k,p_\ell) \,.
\ee

The integrals 
$I^\x  (p_1,p_2,p_3,p_4)  $ 
(with or without numerator factors) 
are functions of the Mandelstam invariants
\begin{align}
s = (p_1 + p_2)^2, \qquad
t = (p_1 + p_4)^2, \qquad
u = (p_1 + p_3)^2 
\end{align}

and hence are invariant under a four element normal subgroup of permutations 
isomorphic to the Klein four-group
\begin{align}
I^\x (p_1, p_2, p_3, p_4) = 
I^\x (p_2, p_1, p_4, p_3) = 
I^\x (p_3, p_4, p_1, p_2) = 
I^\x (p_4, p_3, p_2, p_1) 
\end{align}
or equivalently,
\begin{align}
I_{1234}^\x = 
I_{2143}^\x = 
I_{3412}^\x = 
I_{4312}^\x   \,.
\label{klein}
\end{align}

In the following subsection, we will demonstrate that the SU($N$) gauge group color factors 
are also invariant under the Klein four-group
\begin{align}
c_{1234}^{(x)} = 
c_{2143}^{(x)} = 
c_{3412}^{(x)} = 
c_{4312}^{(x)}   \,.
\label{kleinc}
\end{align}

Hence, using \eqns{klein}{kleinc},
the sum over permutations in \eqn{sumoverperms} 
is reduced to 
\begin{align}
\sum_{S_4} c_{ijkl}^{(x)} I_{ijkl}^\xone
= 
4 \sum_{S_3} c_{1ijk}^{(x)} I_{1ijk}^\xone
\end{align}
where $S_3$ denotes that $ijk$ 
runs over all permutations of $234$.
\para

The precise form
of the $\ell$-loop four-point amplitude
for $\cN=4$ SYM theory is then given by
\cite{Bern:2010tq}
\begin{align}
\cA^\Ell  
~=~
4K (- g_D^2)^{\ell+1}  
 \sum_{x} \sum_{S_3} \lambda^{(x)}  c_{1ijk}^{(x)} I_{1ijk}^\xone
\label{gaugeampprelim}
\end{align}

where $g_D$ is the coupling constant in $D=4-2\de$ dimensions,
and $K$ is a factor common to all loop orders depending
on the helicities of the external states.
For four external gluons, for example, $K$ is given by eq.~(7.4.42) of 
ref.~\cite{Green:2012oqa}.

\subsection{Decomposition in an SU($N$) trace basis}

An alternative representation of the
$\ell$-loop four-point amplitude $\cA^\Ell$
for an SU($N$) gauge theory 
is the decomposition \cite{Bern:1990ux}
\begin{align}
\cA^\Ell  
~=~
\sum_{\lam=1}^6  A^\Ell_{[\lam]}   c_{[\lam]} 
\label{decomposition}
\end{align}

in terms of a six-dimensional basis 
$c_{[\lam]}$ of single and double 
traces
\begin{align}
c_{[1]} &=  
\Tr(T^{a_1} T^{a_2} T^{a_3} T^{a_4}) + \Tr(T^{a_1} T^{a_4} T^{a_3} T^{a_2}),
\qquad
c_{[4]} =  \Tr(T^{a_1} T^{a_3}) \Tr(T^{a_2} T^{a_4}) , 
\nn\\
c_{[2]} &= 
\Tr(T^{a_1} T^{a_2} T^{a_4} T^{a_3}) + \Tr(T^{a_1} T^{a_3} T^{a_4} T^{a_2}),
\qquad
c_{[5]}=  \Tr(T^{a_1} T^{a_4}) \Tr(T^{a_2} T^{a_3})  ,
\label{sixdimbasis} 
\\
c_{[3]} &=  
\Tr(T^{a_1} T^{a_4} T^{a_2} T^{a_3}) + \Tr(T^{a_1} T^{a_3} T^{a_2} T^{a_4}),
\qquad
c_{[6]} =  \Tr(T^{a_1} T^{a_2}) \Tr(T^{a_3} T^{a_4})  \nn
\end{align}

where $a_i$ are the adjoint color indices of the external particles,
and $T^a$ are generators\footnote{Our 
conventions are 
$\Tr(T^a T^b) = \delta^{ab}$ so that 
$[T^a, T^b] = i \sqrt2 f^{abc} T^c$
and 
$f^{abc} = (-i/\sqrt2) \Tr([T^a, T^b] T^c).$}
in the fundamental representation of SU($N$).
All other possible trace terms vanish for SU($N$) since $\Tr(T^a)=0$.
The coefficients $A^\Ell_{[\lam]}$, called color-ordered amplitudes,
are gauge-invariant.
\para

To obtain the color-ordered amplitudes from $\cA^\Ell$,
one replaces the structure constants $\tilde{f}^{abc}$
of an arbitrary color factor $c_{1ijk}$ appearing in \eqn{gaugeampprelim}
with (recalling \eqn{tildef})
\be 
\tilde{f}^{abc} =\Tr( [T^a, T^b]  T^c )
\ee

and then repeatedly utilizes
\begin{align}
\Tr(A T^a) \Tr(B T^a)
&= 
\Tr (AB) - \oneN \Tr(A) \Tr(B) \,,
\nn\\
\Tr(A T^a B T^a)
&= 
\Tr (A) \Tr(B) - \oneN \Tr(A B)
\end{align}

to reduce $c_{1ijk}$ to a linear combination of traces (\ref{sixdimbasis}), 
for example
\be
c_{1234} = \sum_{\lam=1}^6 M_{[\lam]} c_{[\lam]}
\qquad \hbox{or} \qquad  
c_{1234}= ( M_{[1]}, M_{[2]}, M_{[3]}; ~M_{[4]}, M_{[5]}, M_{[6]} )
\label{colorfactordecomp}
\ee

where $M_{[\lam]}$ are polynomials in $N$.
By inspection the elements of the trace basis $c_{[\lam]}$ 
are invariant under the Klein four-group,
and therefore so are any of the color factors $c_{1234}$,
as claimed above in \eqn{kleinc}.
Given \eqn{colorfactordecomp}, 
one can write down the decomposition of 
color factors with permuted legs as
\begin{align}
c_{1243}  &=   ( M_{[2]}, M_{[1]}, M_{[3]}; ~M_{[5]}, M_{[4]}, M_{[6]} ), 
\nn\\[1mm]
c_{1342}  &=   ( M_{[3]}, M_{[1]}, M_{[2]}; ~M_{[6]}, M_{[4]}, M_{[5]} ), 
\nn\\[1mm]
c_{1324}  &=   ( M_{[3]}, M_{[2]}, M_{[1]}; ~M_{[6]}, M_{[5]}, M_{[4]} ), 
\nn\\[1mm]
c_{1423}  &=   ( M_{[2]}, M_{[3]}, M_{[1]}; ~M_{[5]}, M_{[6]}, M_{[4]} ), 
\nn\\[1mm]
c_{1432}  &=   ( M_{[1]}, M_{[3]}, M_{[2]}; ~M_{[4]}, M_{[6]}, M_{[5]} ). 
\label{otherperms}
\end{align}

\subsection{Extended trace basis}

The $\ell$-loop color-ordered amplitudes $A^\Ell_{[\lam]}$ 
can be further decomposed in powers of $N$ as \cite{Bern:1997nh}
\be
A^\Ell_{[\lam]} = 
\begin{cases}
\sum_{k=0}^{\lfloor \frac{\ell}{2}  \rfloor} 
N^{\ell-2k} A^\Elltwok_\lam ,
& \mbox{for } \lam=1,2,3, \\
\sum_{k=0}^{\lfloor \frac{\ell-1}{2}  \rfloor}
N^{\ell-2k-1} A^\Elltwokplus_\lam,
& \mbox{for } \lam=4,5,6 \\
\end{cases}
\ee

where  $A^{(\ell,0)}_\lam$ are leading-order-in-$N$ (planar) amplitudes,
and  $A^\Ellk_\lam$, $k = 1, \cdots, \ell$,  are subleading-order-in-$N$,
yielding $(3\ell+3)$ color-ordered amplitudes at $\ell$ loops.
This suggests an enlargement of the six-dimensional trace basis $c_{[\lam]}$ 
to an extended $(3\ell+3)$-dimensional trace basis $ t^\Ell_\lam $,
defined by 
\begin{align}
t^\Ell_{1+ 6k} &=   N^{\ell - 2k} \, c_{[1]} \,,
\qquad\qquad\qquad t^\Ell_{4+6k} =  N^{\ell - 2k - 1} \, c_{[4]} \,,
\nn \\
t^\Ell_{2+ 6k} &=  N^{\ell-2k} \, c_{[2]} \,,
\qquad\qquad\qquad t^\Ell_{5+6k}  = N^{\ell - 2k - 1}  \, c_{[5]} \,,
\label{extendedbasis}
\\
t^\Ell_{3+ 6k} &=  N^{\ell-2k} \, c_{[3]} \,,
\qquad\qquad\qquad t^\Ell_{6 +6k}=  N^{\ell - 2k - 1} \, c_{[6]} \,.
\nn
\end{align} 

Then \eqn{colorfactordecomp} becomes
\be
c_{1234} = 
 \sum_{\lam =1}^{3\ell+3} m_\lam t^\Ell_\lam
\quad \hbox{or} \quad  
c_{1234}= ( m_1, m_2, m_3;  ~~ m_4, m_5, m_6; 
~~ m_7, m_8, m_9; ~~ \cdots) 
\ee

where $m_\lam$ are integers, 
and \eqn{decomposition} becomes
\be
\cA^\Ell = \sum_{\lam =1}^{3\ell+3}  A^\Ell_\lam t^\Ell_\lam,
\qquad
{\rm where}
\qquad
A^\Ell_{\lam + 6k}=
\begin{cases}
A^\Elltwok_\lam ,    & \lam = 1, 2, 3 \,,\\
A^\Elltwokplus_\lam,  & \lam = 4, 5, 6 \,.
\end{cases}
\label{expandeddecomp}
\ee

The color-ordered amplitudes $A_\lam^\Ell$ are not all independent
but are related by various group-theory constraints \cite{Naculich:2011ep},
which were summarized in ref.~\cite{Naculich:2020clm}.
\para

Since the tree-level color-ordered amplitude is given by \cite{Bern:1998ug}
\be
A_1^\Zero =  (- g_D)^2 {4  K \over st}
\ee

we can rewrite \eqn{gaugeampprelim}
in the form
\begin{align}
\boxed{
\cA^\Ell  
~=~
(- g_D^2)^\ell   s t A_1^\Zero
 \sum_{x} \sum_{S_3} \lambda^{(x)}  c_{1ijk}^{(x)} I_{1ijk}^\xone
}\,.
\label{gaugeamp}
\end{align}

In subsequent sections, explicit expressions for the one-, two-, and three-loop color factors will be used 
to derive the color-ordered amplitudes 
in terms of the scalar integrals $I^\xone$.

\subsection{$\cN=8$ supergravity four-point amplitude}

Using generalized unitarity, 
the $\ell$-loop $\cN=8$ supergravity four-point amplitude
can also be expressed as a linear combination 
of scalar integrals  
\be
I^\xtwo (p_1,p_2,p_3,p_4)  
= (-i)^\ell \int \prod_{i=1}^\ell {d^D \ell_i \over (2\pi)^D} 
{N^\xtwo  \over \prod_{j=1}^{3\ell+1} l_j^2 }
\label{Idef}
\ee

which are analogous to the gauge-theory integrals (\ref{Ixone}), 
but with different numerator factors.
The sum over all permutations of external legs can again be reduced,
using the invariance (\ref{klein}) under the Klein four-group,
to a sum over $S_3$.
\para

The precise form of the $\ell$-loop four-point amplitude
for $\cN=8$ supergravity is then given by
\cite{Bern:1998ug,Bern:2008pv}
\begin{align}
\cM^\Ell  
~=~
16 K \tilde{K} 
(-1)^{\ell+1} \left( \kappa_D \over 2 \right)^{2\ell+2} 
 \sum_{x} \sum_{S_3} \lambda^{(x)}  I_{1ijk}^\xtwo
\label{gravampprelim}
\end{align}
where the combinatorial factors $\lambda^\x$ are the same
as those for the gauge-theory amplitude (\ref{gaugeamp}). 
Since the tree-level four-point gravity amplitude is\cite{Bern:1998ug}
\be
\cM^\Zero = \left(\kappa_D \over 2 \right)^2 {16 K \tilde{K} \over s t u}
\ee

we may rewrite \eqn{gravampprelim} as 
\begin{align}
\boxed{
\cM^\Ell  
~=~
(-1)^{\ell+1}
\left(\kappa_D \over 2 \right)^{2\ell} s t u \cM^\Zero
 \sum_{x} \sum_{S_3} \lambda^{(x)}  I_{1ijk}^\xtwo
} \,.
\label{gravamp}
\end{align}

\section{One-loop relation}
\setcounter{equation}{0}
\label{sec:oneloop}

In this section we review the one-loop $\cN=4$ SYM 
and $\cN=8$ supergravity four-point amplitudes,
and the exact relation between them.
Finally we examine the limiting form of the one-loop 
relation in the Regge limit. 

\subsection{One-loop $\cN=4$ SYM amplitude}

Only the box diagram contributes 
to the one-loop $\cN=4$ SYM four-point 
amplitude \cite{Green:1982sw,Bern:1998ug}
\be
\cA^\One  = - g_D^2 s t A_1^\Zero \sum_{S_3}  \fr{1}{2} c_{1ijk}^\bx   I_{ijkl}^\bx
\label{oneloopsum}
\ee

where the one-loop box color factor is
\be 
c^\bx_{1234} \equiv
 \tf^{e a_1 b} \tf^{b a_2 c} \tf^{c a_3 d} \tf^{d a_4 e}
\ee 

and the box scalar integral is 
(since the numerator factor $N^{ ({\rm box},1)    }=1$)
\be
I_{1ijk}^\bx
\equiv I^\bx (p_1, p_i, p_j, p_k)
=
-i \int {d^D  \ell \over (2 \pi)^D }
{1 \over \ell^2 (\ell-p_1)^2 (\ell-p_1-p_i)^2 (\ell+p_k)^2 } \,.
\label{oneloopintegral}
\ee

One easily ascertains that,
in addition to being invariant under the 
Klein four-group (\ref{klein}),
the one-loop color factor and box integral satisfy
the reflection symmetry
\be
c_{1ijk}^\bx = c_{1kji}^\bx, 
\qquad
\qquad
I_{1ijk}^\bx =I_{1kji}^\bx
\label{boxsymmetries}
\ee
so \eqn{oneloopsum} reduces to
\begin{align}
\cA^\One 
&= - g_D^2 s t A_1^\Zero 
\left[ c_{1234}^\bx I_{1234}^\bx
     + c_{1342}^\bx I_{1342}^\bx
     + c_{1423}^\bx I_{1423}^\bx
\right] \,.
\label{oneloopgauge}
\end{align}

The one-loop color factor can be decomposed into
\begin{align}
c^\bx_{1234}
=  t_1^\One + 2 (t_4^\One + t_5^\One + t_6^\One)  =  (1,0,0;~ 2,2,2)
\label{oneloopcolorfactors}
\end{align}

with other permutations satisfying \eqn{otherperms}.
Consequently, 
the one-loop amplitude can be decomposed 
in the extended trace basis (\ref{expandeddecomp}) 
with color-ordered amplitudes given by 
\begin{align}
A^\One_1
&=
 - g_D^2 s t A_1^\Zero I_{1234}^\bx \,,
\nn\\
A^\One_2
&
=
 - g_D^2 s t A_1^\Zero I_{1342}^\bx \,,
\label{oneloopcolorordered}
\\
A^\One_3
&=
 - g_D^2 s t A_1^\Zero I_{1423}^\bx 
\nn
\end{align}

and the other color-ordered amplitudes satisfying 
$ A^\One_4 = A^\One_5 = A^\One_6 = 2 ( A^\One_1 +A^\One_2 + A^\One_3 ) $.

\subsection{One-loop $\cN=8$ supergravity amplitude and exact relation}

The one-loop
$\cN=8$ supergravity four-point amplitude 
is \cite{Green:1982sw,Bern:1998ug}
\begin{align}
\cM^\One 
&= 
\left( \kappa_D\over 2 \right)^2 
\,
stu
\cM^\Zero 
 \sum_{S_3}  \fr{1}{2} I_{1ijk}^\bx
\end{align}
where again the numerator factor $N^{({\rm box},2)} =1$.
Using the symmetry (\ref{boxsymmetries}), the amplitude reduces  to
\be
\cM^\One 
= 
\left( \kappa_D\over 2 \right)^2 
\,
stu
\cM^\Zero 
\left[ I_{1234}^\bx + I_{1342}^\bx + I_{1423}^\bx \right] \,.
\label{oneloopgrav}
 \ee

One can see the double-copy prescription at work in
\eqn{oneloopgrav};
one simply replaces the
color factors $c^\bx_{ijkl}$ in \eqn{oneloopgauge}
with (constant) kinematic numerators 
to obtain the gravity amplitude.
\para

Comparing \eqn{oneloopgrav} with \eqn{oneloopcolorordered},
one establishes the exact one-loop relation 
\cite{Green:1982sw}
\be
\boxed{
{1 \over   (\kappa_D/2)^2  s t u} {\cM^\One \over \cM^\Zero  }
= 
{- \left[ A_1^{(1)} +A_2^{(1)} + A_3^{(1)}\right]
\over g_D^2 s t A_1^\Zero } 
}
\,.
\label{exactonelooprelation}
\ee

\subsection{One-loop Regge limit}

To determine the behavior of the four-point amplitudes 
in the Regge limit ($|t|\ll s$), we need to examine the
individual integrals that contribute.
The known exact expression for the one-loop box integral \cite{Bern:1993kr}
has the kinematic prefactor 
\be
I_{1ijk}^\bx  \quad\sim\quad {1 \over s_{1i} s_{1k} }
\ee

where 
\be 
s_{ij} \equiv (p_i+p_j)^2 .
\label{defsij}
\ee

In the Regge limit $|t| \ll s$
(so that $u=-s-t \sim -s$),
one thus has 
$I_{1234}^\bx \sim  1/(st)$ and 
$I_{1423}^\bx \sim 1/(ut) \sim -1/(st)$
whereas $I_{1342}^\bx \sim 1/(su) \sim - 1/s^2$, 
so that 
\be 
I_{1234}^\bx, \quad I_{1423}^\bx\ \quad \gg \quad  I_{1342}^\bx \,.
\label{oneloopordering}
\ee

From \eqns{oneloopcolorordered}{oneloopordering} 
one then has that the $\cN=4$ SYM color-ordered amplitudes obey
\be 
A_1^{(1)}, \quad A_3^{(1)} 
\quad \gg \quad 
A_2^{(1)} 
\ee 

so the exact one-loop SYM/supergravity relation (\ref{exactonelooprelation}) 
reduces in the Regge limit to 
\be
{1 \over   (\kappa_D/2)^2  s t u} {\cM^\One \over \cM^\Zero  }
\regge
{- \left[ A_1^{(1)}  + A_3^{(1)}\right]
\over g_D^2 s t A_1^\Zero }
\label{onelooprelationbis}
\ee

in agreement with \eqn{onelooprelation}.
Moreover, \eqn{oneloopordering}
implies that the one-loop supergravity amplitude (\ref{oneloopgrav}) 
reduces in the Regge limit to 
\begin{align}
\cM^\One 
&\regge
\left( \kappa_D\over 2 \right)^2 
\,
stu
\cM^\Zero 
\left[ I_{1234}^\bx  + I_{1423}^\bx \right] 
\label{oneloopladders}
\end{align}
the sum of a box and a ``crossed-box'' diagram.
We will revisit this in sec.~\ref{sec:ladders}.

\section{Two-loop relation}
\setcounter{equation}{0}
\label{sec:twoloop}

In this section we review the two-loop $\cN=4$ SYM 
and $\cN=8$ supergravity four-point amplitudes \cite{Bern:1998ug}.
We will then re-establish the known exact relation \cite{Naculich:2008ys}
between them,
as well as its limiting form in the Regge limit. 

\subsection{Two-loop $\cN=4$ SYM amplitude}

\begin{figure}[b]
\begin{center}
\includegraphics[width=4.0cm]{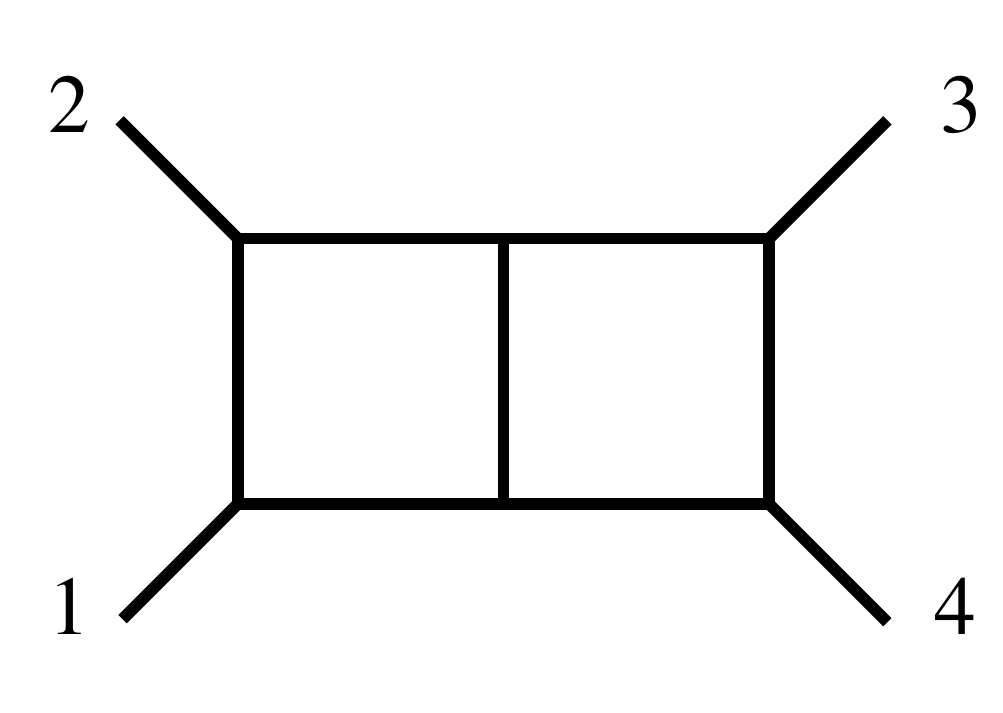}
\hskip2cm
\includegraphics[width=4.0cm]{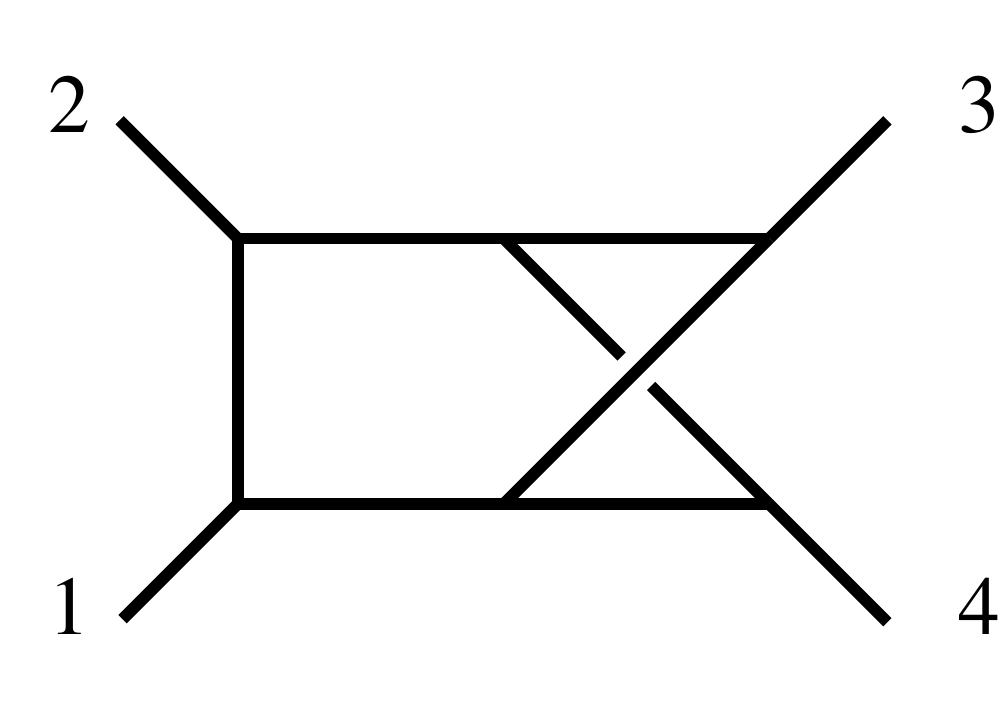}
\caption{Two-loop diagrams P and NP}
\label{fig:twoloop}
\end{center}
\end{figure}

The two-loop $\cN=4$ SYM four-point amplitude 
is given by \cite{Bern:1997nh,Bern:1998ug}
\begin{align}
\cA^\Two 
=  g_D^4 ~s t A_1^\Zero 
\sum_{S_3} \Big[ c_{1ijk}^\P I_{1ijk}^\Pone  
+ c_{1ijk}^\NP I_{1ijk}^\NPone  \Big]
\label{twoloopyangmills}
\end{align}

where $P$ and $NP$ denote 
the planar double box and non-planar diagrams
shown in fig.~\ref{fig:twoloop},
and the associated two-loop color factors are
\begin{align}
c^\P_{1234}
&= \tf^{e a_1 b} \tf^{b a_2 c} \tf^{cgd} \tf^{dfe} \tf^{g a_3 h} \tf^{h a_4 f} \,,
\nn \\
c^\NP_{1234}
&= \tf^{e a_1 b} \tf^{b a_2 c} \tf^{cgd} \tf^{hfe} \tf^{g a_3 h} \tf^{d a_4 f} \,.
\end{align}

The two-loop integrals {\it without numerator factors} are 
\begin{align}
I^\Pzero_{1ijk}
& = 
-  \int {d^D \ell_1 \over (2\pi)^D} {d^D \ell_2 \over (2\pi)^D}
{ 1 \over \ell_1^2 \, (\ell_1 + \ell_2)^2 \ell_2^2 \, (\ell_1 - p_1)^2 \,(\ell_1 - p_1 - p_i)^2 \,
        (\ell_2-p_k)^2 \, (\ell_2 - p_j - p_k)^2 } \,,
\nn \\[1mm]
\\
I^\NPzero_{1ijk} 
&= 
-  \int {d^D \ell_1 \over (2\pi)^D} {d^D \ell_2 \over (2\pi)^D}
{1  \over \ell_1^2\,(\ell_1+\ell_2)^2\, \ell_2^2 \, (\ell_1-p_i)^2  \,(\ell_1+\ell_2+p_1)^2\,
  (\ell_2-p_j)^2 \, (\ell_2-p_j-p_k)^2} \,.
\nn 
\end{align}

Including the two-loop gauge-theory numerator factors,
we have 
\be
I^\Pone_{1ijk} =
s_{1i}
I^\Pzero_{1ijk} ,
\qquad\qquad
I^\NPone_{1ijk} =
s_{1i}
I^\NPzero_{1ijk}  \,.
\label{onezero}
\ee

The non-planar color factor and integral have the additional symmetries 
\be 
 c^\NP_{1ijk} = c^\NP_{1ikj},
\qquad\qquad
I_{1ijk}^\NPone = I_{1ikj}^\NPone 
\label{twoloopnonplanarsymmetry}
\ee 

so \eqn{twoloopyangmills} reduces to 
\begin{align} 
\cA^\Two 
=  g_D^4 ~s t A_1^\Zero 
 \Big[ 
& c_{1234}^\P I_{1234}^\Pone 
+c_{1243}^\P I_{1243}^\Pone 
+c_{1342}^\P I_{1342}^\Pone 
+c_{1324}^\P I_{1324}^\Pone 
+c_{1423}^\P I_{1423}^\Pone 
+c_{1432}^\P I_{1432}^\Pone 
\nn\\
&+
2 c_{1234}^\NP I_{1234}^\NPone 
+2c_{1342}^\NP I_{1342}^\NPone 
+2c_{1423}^\NP I_{1423}^\NPone 
\Big] \,.
\label{twoloopamplitude}
\end{align}

The two-loop color factors may be decomposed in the 
extended trace basis as 
\be
\begin{array}{rrrrrrrrr}
c^\P_{1234}
=
(1,&0,&0; & 0,&0,&6;&2,&2,&-4) \,, \\[1mm]
c^\NP_{1234}
=
(0,&0,&0;&-2,&-2,&4;& 2,&2,&-4)  \,.
\end{array}
\label{twoloopextendedtrace}
\ee

Plugging \eqn{twoloopextendedtrace} together with
the other permutations obtained via \eqn{otherperms}
into \eqn{twoloopamplitude},
we compute the color-ordered amplitudes \cite{Bern:1997nh}
\begin{align}
A_1^\Two
&= 
  g_D^4 ~s t A_1^\Zero 
\Big[
 I_{1234}^\Pone 
+I_{1432}^\Pone 
\Big],
\nn\\
A_2^\Two
&= 
  g_D^4 ~s t A_1^\Zero 
\Big[
 I_{1243}^\Pone 
+I_{1342}^\Pone 
\Big],
\nn\\
A_3^\Two
&= 
  g_D^4 ~s t A_1^\Zero 
\Big[
 I_{1324}^\Pone 
+I_{1423}^\Pone 
\Big],
\nn\\
A_4^\Two
&= 
 2 g_D^4 ~s t A_1^\Zero 
\Big[
 3I_{1342}^\Pone 
+3I_{1324}^\Pone 
-2I_{1234}^\NPone 
+4I_{1342}^\NPone 
-2I_{1423}^\NPone 
\Big],
\nn\\
A_5^\Two
&= 
 2 g_D^4 ~s t A_1^\Zero 
\Big[
 3I_{1423}^\Pone 
+3I_{1432}^\Pone 
-2I_{1234}^\NPone 
-2I_{1342}^\NPone 
+4I_{1423}^\NPone 
\Big],
\label{twoloopcolorordered}
\\
A_6^\Two
&= 
 2 g_D^4 ~s t A_1^\Zero 
\Big[
 3I_{1234}^\Pone 
+3I_{1243}^\Pone 
+4I_{1234}^\NPone 
-2I_{1342}^\NPone 
-2I_{1423}^\NPone 
\Big],
\nn\\
A_7^\Two
&= 
 2 g_D^4 ~s t A_1^\Zero 
\Big[
 I_{1234}^\Pone 
+I_{1243}^\Pone 
-2I_{1342}^\Pone 
-2I_{1324}^\Pone 
+I_{1423}^\Pone 
+I_{1432}^\Pone 
+2I_{1234}^\NPone 
-4I_{1342}^\NPone 
+2I_{1423}^\NPone 
\Big],
\nn\\
A_8^\Two
&= 
 2 g_D^4 ~s t A_1^\Zero 
\Big[
 I_{1234}^\Pone 
+I_{1243}^\Pone 
+I_{1342}^\Pone 
+I_{1324}^\Pone 
-2I_{1423}^\Pone 
-2I_{1432}^\Pone 
+2I_{1234}^\NPone 
+2I_{1342}^\NPone 
-4I_{1423}^\NPone 
\Big],
\nn\\
A_9^\Two
&= 
 2 g_D^4 ~s t A_1^\Zero 
\Big[
-2 I_{1234}^\Pone 
-2I_{1243}^\Pone 
+ I_{1342}^\Pone 
+ I_{1324}^\Pone 
+I_{1423}^\Pone 
+I_{1432}^\Pone 
-4I_{1234}^\NPone 
+2I_{1342}^\NPone 
+2I_{1423}^\NPone 
\Big].
\nn
\end{align}

From these, using $s+t+u=0$, 
we easily obtain the linear combination
\begin{align}
{ 
u A_7^\Two  + t A_8^\Two  + s A_9^\Two  \over
 g_D^4  s t A_1^\Zero  }
=
-
6 \Bigl[
&  s (I_{1234}^\Pone 
+  I_{1243}^\Pone )
+ u (I_{1342}^\Pone 
+  I_{1324}^\Pone)
+ t (I_{1423}^\Pone 
+  I_{1432}^\Pone)
\nn\\
& 
+2s I_{1234}^\NPone 
+2u I_{1342}^\NPone 
+2t I_{1423}^\NPone 
\Big]
\label{sumtwoloop}
\end{align}
which will be used below.

\subsection{Two-loop $\cN=8$ supergravity amplitude and exact relation}

The two-loop $\cN=8$ supergravity four-point amplitude 
is given by
\cite{Bern:1998ug}
\begin{align}
\cM^\Two 
&=  
-  \left( \kappa_D \over 2\right)^4   
 ~ s t u \cM^\Zero 
\sum_{S_3} \Big[ I_{1ijk}^\Ptwo  + I_{1ijk}^\NPtwo  \Big]
\label{twoloopgravpre}
\end{align}

where the two-loop gravity integrals 
include the numerator factors
\be
I^\Ptwo_{1ijk} =
s_{1i}^2
I^\Pzero_{1ijk} ,
\qquad\qquad
I^\NPtwo_{1ijk} =
s_{1i}^2 
I^\NPzero_{1ijk}  \,.
\label{twozero}
\ee

Using the symmetry $ I_{1ijk}^\NPtwo = I_{1ikj}^\NPtwo$, 
we may write  \eqn{twoloopgravpre} as
\begin{align}
\cM^\Two 
=
-  \left( \kappa_D \over 2\right)^4   
 ~ s t u \cM^\Zero 
 \Big[ 
& 
 I_{1234}^\Ptwo 
+I_{1243}^\Ptwo 
+I_{1342}^\Ptwo 
+I_{1324}^\Ptwo 
+I_{1423}^\Ptwo 
+I_{1432}^\Ptwo 
\nn\\
&\hskip20mm
+
 2I_{1234}^\NPtwo 
+2I_{1342}^\NPtwo 
+2I_{1423}^\NPtwo 
\Big] \,.
\label{twoloopgrav}
\end{align}

From \eqns{onezero}{twozero} we have 
$
I^\Ptwo_{1ijk} =
s_{1i}
I^\Pone_{1ijk} 
$ 
and  
$
I^\NPtwo_{1ijk} =
s_{1i}
I^\NPone_{1ijk} 
$ 
so that 
\begin{align}
\cM^\Two 
=
-  \left( \kappa_D \over 2\right)^4   
 ~ s t u \cM^\Zero 
 \Big[ 
&  s (I_{1234}^\Pone 
+  I_{1243}^\Pone )
+ u (I_{1342}^\Pone 
+  I_{1324}^\Pone)
+ t (I_{1423}^\Pone 
+  I_{1432}^\Pone)
\nn\\
& 
+2s I_{1234}^\NPone 
+2u I_{1342}^\NPone 
+2t I_{1423}^\NPone 
\Big] \,.
\label{twoloopgravagain}
\end{align}

This expression can be understood in terms of the double copy 
by replacing the color factors of \eqn{twoloopamplitude}
with kinematic numerators; 
see ref.~\cite{BoucherVeronneau:2011qv} for details.
\para

Comparing \eqns{sumtwoloop}{twoloopgravagain},
one obtains the exact two-loop 
relation \cite{Naculich:2008ys}
\be
\boxed{
 {1 \over ( \kappa_D/2)^4    ~s t u }
 {\cM^\Two \over \cM^\Zero }
= 
{\fr{1}{6} \left[u A_7^\Two  + t A_8^\Two  + s A_9^\Two  \right] \over
 g_D^4  s t A_1^\Zero  }
} \,.
\label{exacttwolooprelation}
\ee

\subsection{Two-loop Regge limit}

To determine the Regge limit of the two-loop amplitude,
we examine the kinematic prefactors of the contributing integrals.
The known explicit expression for the planar double box integral 
\cite{Smirnov:1999gc}
has prefactor 
\be
I_{1ijk}^\Pzero \quad\sim\quad {1 \over s_{1i}^2 s_{1k} }  \,.
\label{Pzero}
\ee

The known explicit expression for the nonplanar integral 
\cite{Tausk:1999vh}
has separate contributions with prefactors
\be
I_{1ijk}^\NPzero \quad\sim\quad 
{1 \over s_{1i}^2 s_{1j} } ~,~
{1 \over s_{1i}^2 s_{1k} }  
\label{NPzero}
\ee

which ensures that the symmetry 
$I_{1ijk}^\NPzero= I_{1ikj}^\NPzero$
is respected.
The gauge-theory integrals satisfy \eqn{onezero} so
\be
I_{1ijk}^\Pone \quad\sim\quad {1 \over s_{1i} s_{1k} } ,
\qquad\qquad
I_{1ijk}^\NPone 
\quad\sim\quad {1 \over s_{1i} s_{1j} } ~,~ {1 \over s_{1i} s_{1k} }  \,.
\label{twoloopSYMbehavior}
\ee

Thus in the Regge limit, one has 
$I_{1234}^\Pone, I_{1324}^\Pone, I_{1432}^\Pone, I_{1423}^\Pone \sim 1/(st)$ 
and 
$I_{1243}^\Pone, I_{1342}^\Pone  \sim 1/s^2$,
similar to the one-loop box integral,
while all permutations of the nonplanar integral $I^\NPone$
go as $1/(st)$.
All in all, we have
\be
I_{1234}^\Pone, I_{1324}^\Pone, I_{1432}^\Pone, I_{1423}^\Pone, 
I_{1234}^\NPone, 
I_{1342}^\NPone, 
I_{1423}^\NPone
\qquad\gg\qquad 
I_{1243}^\Pone, I_{1342}^\Pone  \,.
\label{twoloopgaugeordering}
\ee

Given \eqn{twoloopgaugeordering},
we see from \eqn{twoloopcolorordered} that 
$A_7^\Two$, $A_8^\Two$,  and $A_9^\Two$
are all comparable in the Regge limit, 
so that the exact relation (\ref{exacttwolooprelation})
reduces in this limit  to 
\be
 {1 \over ( \kappa_D/2)^4    ~s t u }
 {\cM^\Two \over \cM^\Zero }
\regge
{- \fr{1}{6} s \left[ A_7^\Two  - A_9^\Two  \right] \over
 g_D^4  s t A_1^\Zero  }
\label{twolooprelationbis}
\ee
in agreement with \eqn{twolooprelation}.
\para

It is instructive to explicitly express 
\eqn{twolooprelationbis}
in terms of the contributing integrals. 
Using \eqn{twoloopcolorordered}
together with \eqn{twoloopgaugeordering}
we have in the Regge limit
\begin{align}
A_7^\Two - A_9^\Two
&\regge
 6 g_D^4 ~s t A_1^\Zero 
\Bigl[
 I_{1234}^\Pone 
-I_{1324}^\Pone 
+2I_{1234}^\NPone
-2I_{1342}^\NPone
\Big]
\nn\\
&\regge
 g_D^4 ~s t A_1^\Zero 
(6 s) 
\Bigl[
 I_{1234}^\Pzero 
+ I_{1324}^\Pzero 
+2I_{1234}^\NPzero
+2I_{1342}^\NPzero
\Big] \,.
\label{rhsregge}
\end{align}

The gravity integrals satisfy \eqn{twozero} so \eqns{Pzero}{NPzero} imply
\be
I_{1ijk}^\Ptwo \quad\sim \quad{1 \over s_{1k} }  , 
\qquad 
\qquad 
I_{1ijk}^\NPtwo 
\quad\sim\quad {1 \over s_{1j} } ~,~ {1 \over s_{1k} } 
\ee

and thus
$I_{1234}^\Ptwo, I_{1324}^\Ptwo, 
I_{1234}^\NPtwo, I_{1342}^\NPtwo$
go as $1/t$, 
while 
$
I_{1243}^\Ptwo, I_{1342}^\Ptwo,
I_{1432}^\Ptwo, I_{1423}^\Ptwo, 
I_{1234}^\NPtwo
$
go as $1/s$ in the Regge limit.
All in all,
\be
I_{1234}^\Ptwo, I_{1324}^\Ptwo, 
I_{1234}^\NPtwo, 
I_{1342}^\NPtwo
\qquad\gg\qquad 
I_{1243}^\Ptwo, I_{1342}^\Ptwo,
I_{1432}^\Ptwo, I_{1423}^\Ptwo, 
I_{1423}^\NPtwo \,.
\label{twoloopgravityordering}
\ee

Thus the two-loop supergravity amplitude 
(\ref{twoloopgrav}) reduces in the Regge limit to 
\begin{align}
\cM^\Two 
&\regge
-  \left( \kappa_D \over 2\right)^4   
 ~ s t u \cM^\Zero 
 \Big[ 
 I_{1234}^\Ptwo 
+I_{1324}^\Ptwo 
+2 I_{1234}^\NPtwo 
+2 I_{1342}^\NPtwo 
\Big]
\nn\\
&\regge
-  \left( \kappa_D \over 2\right)^4   
 ~ s t u \cM^\Zero 
s^2  \Big[ 
 I_{1234}^\Pzero
+I_{1324}^\Pzero
+2 I_{1234}^\NPzero
+2 I_{1342}^\NPzero
\Big] \,.
\label{twoloopladders}
\end{align}

In sec.~\ref{sec:ladders}, we will see that this is just the
sum over all ladder and crossed-ladder diagrams.
\para

Comparing \eqns{rhsregge}{twoloopladders}, we can see explicitly that 
\eqn{twolooprelationbis} is satisfied.

\section{Three-loop relation}
\setcounter{equation}{0}
\label{sec:threeloop}

In this section we present 
the expressions obtained in refs.~\cite{Bern:2007hh,Bern:2008pv}
for the three-loop $\cN=4$ SYM and $\cN=8$ supergravity four-point amplitudes 
in terms of a sum of scalar integrals. 
Unlike the one- and two-loop cases, 
we will not be able to establish an exact relation 
between the three-loop amplitudes (but see subsec.~\ref{sec:threeloop}.4).
However, by examining the Regge limits of the relevant integrals
we will verify the Regge limit relation
\be
\boxed{
 {1 \over ( \kappa_D/2)^6    ~s t u }
 {\cM^\Three \over \cM^\Zero }
\regge
\frac{s^2}{12} 
~{ \left[
  4 \left(  A_1^\Three + A_3^\Three \right) 
- \left( A_4^\Three +A_6^\Three + A_7^\Three +A_9^\Three \right)
\right]
 \over
g_D^6  s t A_1^\Zero  }
}
\label{threelooprelationbis}
\ee
that was conjectured in ref.~\cite{Naculich:2020clm}.

\subsection{Three-loop $\cN=4$ SYM amplitude}

\begin{figure}[p]
\begin{center}
\includegraphics[width=13.5cm]{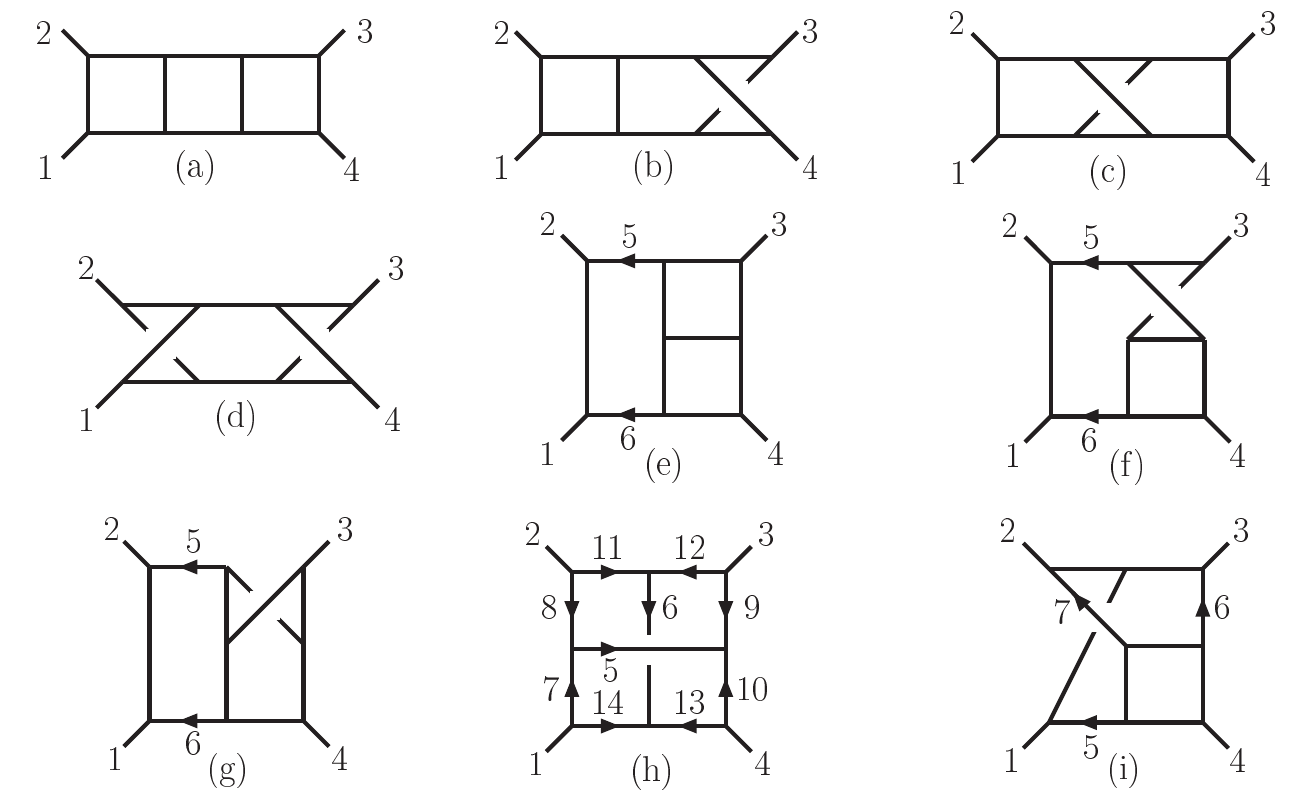}
\caption{Three-loop diagrams (courtesy of ref.~\cite{Bern:2008pv}).}
\label{IntegralsThreeLoopFigure} 
\end{center}
\end{figure}

The three-loop $\cN=4$ SYM four-point amplitude
is given by \cite{Bern:2007hh,Bern:2008pv}
\begin{align}
 \cA^\Three
=
-g_D^6 s t A_1^\Zero 
\sum_{S_3} \Big[
& c_{1ijk}^{(a)} I_{1ijk}^{(a,1)}
+
c_{1ijk}^{(b)} I_{1ijk}^{(b,1)}
+
\fr{1}{2}  c_{1ijk}^{(c)} I_{1ijk}^{(c,1)}
+
\fr{1}{4}  c_{1ijk}^{(d)} I_{1ijk}^{(d,1)}
\label{threeloopyangmills}
\\
& 
+
2 c_{1ijk}^{(e)} I_{1ijk}^{(e,1)}
+
2 c_{1ijk}^{(f)} I_{1ijk}^{(f,1)}
+
4 c_{1ijk}^{(g)} I_{1ijk}^{(g,1)}
+
\fr{1}{2}  c_{1ijk}^{(h)} I_{1ijk}^{(h,1)}
+
2 c_{1ijk}^{(i)} I_{1ijk}^{(i,1)}
\Big]
\nn
\end{align}

where $c^\x$ and $I^\xone$  are color factors and scalar 
integrals associated with the nine diagrams shown 
in fig.~\ref{IntegralsThreeLoopFigure}.
The $\cN=4$ SYM numerator factors 
$N^\xone$ appearing in the integrals $I^\xone$
are given in ref.~\cite{Bern:2008pv}
and reproduced in table 1,
with the invariants appearing in the table defined as 
\be
\tau_{ij} = 2 p_i\cdot l_j\,,
\qquad
 s_{ij} 
= \Bigg\{
\begin{array}{ll}
(p_i +p_j)^2, &  \hskip 1cm i,j \le 4; \\
(p_i +l_j)^2, &  \hskip 1cm i \le 4<j;  \\
(l_i +l_j)^2, &  \hskip 1cm  4<i,j.
\end{array}
\ee

The three-loop color factors may be decomposed in the 
extended trace basis as

\be
\begin{array}{rrrrrrrrrrrr}
 c_{1234}^{(a)} =
 (1 ,& 0 ,& 0 ;& 0 ,& 0 ,& 14 ;& 2 ,& 2 ,& 0 ;& 8 ,& 8 ,& 8 ),\\[1mm]
 c_{1234}^{(b)} =
 (0 ,& 0 ,& 0 ;& 0 ,& 0 ,& 8 ;& 0 ,& 0 ,& 4 ;& 8 ,& 8 ,& 8 ),\\[1mm]
 c_{1234}^{(c)} =
 (0 ,& 0 ,& 0 ;& 0 ,& 0 ,& 8 ;& 0 ,& 0 ,& 4 ;& 8 ,& 8 ,& 8 ),\\[1mm]
 c_{1234}^{(d)} =
 (0 ,& 0 ,& 0 ;& 2 ,& 2 ,& 4 ;& -2 ,& -2 ,& 8 ;& 8 ,& 8 ,& 8 ),\\[1mm]
 c_{1234}^{(e)} =
 (1 ,& 0 ,& 0 ;& 0 ,& 0 ,& 2 ;& 8 ,& -4 ,& -6 ;& -4, & -4 ,& -4 ),\\[1mm]
 c_{1234}^{(f)} =
 (0 ,& 0 ,& 0 ;& -2 ,& -2 ,& 0 ;& 8 ,& -4 ,& -6 ;& -4 ,& -4 ,& -4 ),\\[1mm]
 c_{1234}^{(g)} =
 (0 ,& 0 ,& 0 ;& 0 ,& 0 ,& -4 ;& 6 ,& -6 ,& -2 ;& -4 ,& -4 ,& -4 ),\\[1mm]
 c_{1234}^{(h)} =
 (0 ,& 0 ,& 0 ;& 0 ,& 2 ,& 2 ;& -6 ,& 4 ,& 4 ;& 4 ,& 4 ,& 4 ),\\[1mm]
 c_{1234}^{(i)} =
 (0 ,& 0 ,& 0 ;& 0 ,& -2 ,& 2 ;& 0 ,& 2 ,& -2 ;& 0 ,& 0 ,& 0 ).
\end{array}
\label{threeloopextendedtrace}
\ee

\begin{table}[p]
\begin{center}
\begin{tabular}{||c|c|}
\hline
$
\vphantom{\Big|}
I^\xone $ & $N^\xone $ \\
\hline
\hline
(a)--(d) &  $
\vphantom{\Bigr|}
s_{12}^2
$
  \\
\hline
(e)--(g) &  $
\vphantom{\Bigr|}
s_{12} \,s_{46}
$  \\
\hline
(h)&  $\;
\vphantom{\Bigr|}
s_{12} (\tau_{26} + \tau_{36}) +
       s_{14} (\tau_{15} + \tau_{25}) +
       s_{12} s_{14}
\;$
\\
\hline
(i) & $\;
\vphantom{\Bigr|}
s_{12} s_{45} - s_{14} s_{46} -  {1\over 3} (s_{12} - s_{14}) l_7^2
 \;$  \\
\hline
\end{tabular}
\caption{
$\cN=4$ SYM numerator factors  (from ref.~\cite{Bern:2008pv})
\label{NumeratorYMTable} }
\end{center}
\end{table}

\begin{table}[p]
\begin{center}
\begin{tabular}{||c|c||}
\hline
$
\vphantom{\Big|}
I^\xtwo $ & $N^\xtwo $ \\
\hline
\hline
(a)--(d)
& $
\vphantom{\Bigr|}
 [s_{12}^2]^2
$ \\
\hline
(e)--(g) & $\vphantom{\Bigr|}
 [s_{12} \, s_{46}]^2
$  \\
\hline
(h) & $\;
\vphantom{\Bigr|}
(s_{12} s_{89} + s_{14} s_{11,14} - s_{12} s_{14})^2
          - s_{12}^2 (2 (s_{89} - s_{14}) + l_6^2 ) l_6^2
          - s_{14}^2 (2 (s_{11,14} - s_{12}) + l_5^2)  l_5^2
         \; $  \\
    &  $ \; \null
\vphantom{\Bigr|}
      - s_{12}^2 (2 l_8^2 l_{10}^2 + 2 l_7^2 l_9^2 + l_8^2 l_7^2
        + l_9^2 l_{10}^2)
       \vphantom{\bigl(A^A_A\bigr) }
 - s_{14}^2 (2 l_{11}^2 l_{13}^2 + 2 l_{12}^2 l_{14}^2 +
            l_{11}^2 l_{12}^2 + l_{13}^2 l_{14}^2 )
           + 2 s_{12} s_{14} l_5^2 l_6^2\;
         $
           \\
\hline
(i) &  $
\vphantom{\Bigr|}
(s_{12} s_{45} - s_{14} s_{46} )^2
  \vphantom{\bigl|_{A_A}} \null - ( s_{12}^2  s_{45} + s_{14}^2 s_{46}
      + {1\over3} s_{12} s_{13} s_{14}) l_7^2
 $ \\
\hline
\end{tabular}
\caption{$\cN=8$ supergravity numerator factors (from ref.~\cite{Bern:2008pv})}
\end{center}
\end{table}

To obtain the three-loop color-ordered amplitudes $A^\Three_\lam$,
one simply substitutes \eqn{threeloopextendedtrace},
together with other permutations obtained via \eqn{otherperms},
into \eqn{threeloopyangmills} 
and reads off the coefficient of each $t_\lam^\Three$
as a linear combination of the integrals $I^\xone$.
From these expressions, which we will not reproduce here, 
one can compute
the particular linear combination
appearing on the r.h.s. of the relation (\ref{threelooprelationbis})

\begin{align} 
&
\hskip-25mm
 { 4 \left(  A_1^\Three + A_3^\Three \right) 
- \left( A_4^\Three +A_6^\Three + A_7^\Three +A_9^\Three \right)
 \over
g_D^6  s t A_1^\Zero  }
\nn\\
= 
12 \Big[  
& I_{1234}^{(a,1)} 
+\fr{4}{3} I_{1243}^{(a,1)}
+ I_{1324}^{(a,1)}
+\fr{4}{3} I_{1342}^{(a,1)}
+ 2 I_{1234}^{(b,1)} 
+2 I_{1342}^{(b,1)}
+ \fr{1}{2}  \left(  
2 I_{1234}^{(c,1)} 
+2 I_{1342}^{(c,1)}
\right)
\nn\\ &
+ \fr{1}{4}  \left(  
 2 I_{1234}^{(d,1)} 
+2 I_{1342}^{(d,1)}
\right)
+ 2  \left(  
-\fr{2}{3} I_{1243}^{(e,1)}
-\fr{2}{3} I_{1342}^{(e,1)}
\right)
+2  \left( 
- I_{1243}^{(f,1)}
- I_{1342}^{(f,1)}
\right)
\nn\\ &
+4 \left(
-  I_{1243}^{(g,1)}
- I_{1342}^{(g,1)}
\right)
+\fr{1}{2}  \left(
  I_{1243}^{(h,1)}
+ I_{1342}^{(h,1)}
\right)
\Big]
\label{threelooprhs}
\end{align}

where we have used 
\be
I_{1ijk}^{(b,1)} = I_{1ikj}^{(b,1)}, \qquad\qquad
I_{1ijk}^{(c,1)} = I_{1ikj}^{(c,1)}, \qquad\qquad
I_{1ijk}^{(d,1)} = I_{1ikj}^{(d,1)} 
\label{threeloopnonplanarsymmetry}
\ee

which are manifestly satisfied for the nonplanar diagrams 
$b$, $c$, and $d$.

\subsection{Three-loop $\cN=8$ supergravity amplitude}

The three-loop $\cN=8$ supergravity four-point amplitude
is given by \cite{Bern:2007hh,Bern:2008pv}
\begin{align}
 \cM^\Three
=
 \left( \kappa_D \over 2 \right)^6  
s t u  \cM^\Zero 
\sum_{S_3} \Big[
 &
I_{1ijk}^{(a,2)}
+I_{1ijk}^{(b,2)}
+\fr{1}{2}  I_{1ijk}^{(c,2)}
+\fr{1}{4}  I_{1ijk}^{(d,2)}
\nn\\
&+2I_{1ijk}^{(e,2)}
+2I_{1ijk}^{(f,2)}
+4I_{1ijk}^{(g,2)}
+\fr{1}{2} I_{1ijk}^{(h,2)}
+2 I_{1ijk}^{(i,2)}
\Big]
\label{threeloopgrav}
\end{align}

where $I^\xtwo$ are scalar integrals associated 
with the nine diagrams shown 
in fig.~\ref{IntegralsThreeLoopFigure}.
The $\cN=8$ SYM numerator factors 
$N^\xtwo$ appearing in the integrals $I^\xtwo$
are given in ref.~\cite{Bern:2008pv}
and reproduced in table 2.
In general there are no additional symmetries
among the integrals except for the 
three nonplanar diagrams $b$, $c$, and $d$,
which manifestly satisfy 
\be
I_{1ijk}^{(b,2)} = I_{1ikj}^{(b,2)}, \qquad\qquad
I_{1ijk}^{(c,2)} = I_{1ikj}^{(c,2)}, \qquad\qquad
I_{1ijk}^{(d,2)} = I_{1ikj}^{(d,2)}. 
\label{threeloopnonplanarsymmetrygrav}
\ee

\subsection{Three-loop Regge limit}

We now examine the Regge limit of the integrals
appearing in \eqns{threelooprhs}{threeloopgrav}.
In particular, we will find that 
\begin{itemize}
\item 
the leading integrals in the SYM amplitude go as $1/(st)$, 
\item
the leading integrals in the supergravity amplitude go as $s/t$,
and
\item
the other integrals are suppressed by one or more factors of $t/s$.
\end{itemize}
This will lead to simplified expressions for the four-point amplitudes
in the Regge limit.
\para

We begin with the two planar diagrams, $a$ and $e$, 
for which explicit expressions 
for the gauge-theory integrals 
are known \cite{Bern:2005iz},
from which we see that they have kinematic prefactors
\be
I^{(a,1)}_{1ijk} \sim {1 \over s_{1i} s_{1k}} 
,\qquad
\qquad
I^{(e,1)}_{1ijk} \sim {1 \over s_{1i} s_{1k}} \,.
\label{diagramae}
\ee

This implies that, 
in the Regge limit,
\be
I_{1234}^{(a,1)}, I_{1324}^{(a,1)} 
\qquad\gg\qquad
I_{1243}^{(a,1)}, I_{1342}^{(a,1)}, I_{1243}^{(e,1)}, I_{1342}^{(e,1)} 
\label{limita}
\ee
so that the four gauge-theory integrals on the r.h.s. of \eqn{limita}, 
which go as $1/s^2$,  
can be neglected 
relative to the two on the l.h.s., which go as $1/(st)$.
(We only consider those integrals appearing in \eqn{threelooprhs}.)
\para

The numerator factor for the gravity integral 
$I^{(a,2)}_{1234} $
is $s_{12}^4$,
compared to $s_{12}^2$ for gauge theory, 
so that
$I_{1ijk}^{(a,2)} = s_{1i}^2 I_{1ijk}^{(a,1)}$.
Hence from \eqn{diagramae}, we have
\be
I_{1ijk}^{(a,2)} \quad\sim \quad{s_{1i} \over s_{1k} }  \,.
\ee
Therefore, in the Regge limit
\be
I_{1234}^{(a,2)}, I_{1324}^{(a,2)} 
\qquad\gg\qquad
I_{1243}^{(a,2)}, I_{1342}^{(a,2)}
\qquad\gg\qquad
I_{1432}^{(a,2)}, I_{1423}^{(a,2)} 
\label{gravitylimita}
\ee
so only the two integrals on the l.h.s. of \eqn{gravitylimita},
which go as $s/t$,
contribute to \eqn{threeloopgrav} for $\cM^\Three$ 
to leading order in the Regge limit.
\para

Next, we expect that,
like the two-loop nonplanar integrals (\ref{twoloopSYMbehavior}),
the three-loop nonplanar gauge-theory integrals corresponding
to $b$, $c$, $d$ have separate contributions with kinematic prefactors
\be
I^{(b,1)}_{1ijk} , 
I^{(c,1)}_{1ijk} , 
I^{(d,1)}_{1ijk} 
\quad\sim\quad {1 \over s_{1i} s_{1j} } ~,~ {1 \over s_{1i} s_{1k} } 
\label{limitbcd}
\ee

to ensure the symmetry (\ref{threeloopnonplanarsymmetry}).
Hence the contributions 
from these integrals appearing in \eqn{threelooprhs}
all go as $1/(st)$ in the Regge limit.
The corresponding gravity integrals satisfy 
$I_{1ijk}^\xtwo = s_{1i}^2 I_{1ijk}^\xone$,
so from \eqn{limitbcd}, we have
\be
I^{(b,2)}_{1ijk} , 
I^{(c,2)}_{1ijk} , 
I^{(d,2)}_{1ijk} 
\quad\sim\quad {s_{1i} \over s_{1j} } ~,~ {s_{1i} \over s_{1k} }  \,.
\ee

Hence in the Regge limit the gravity integrals satisfy
\be
I^\xtwo_{1234} , 
I^\xtwo_{1342} 
\qquad\gg\qquad
I^\xtwo_{1423} 
\qquad \hbox{for } x=b, c, d
\label{gravitylimitbcd}
\ee

so only the two integrals on the l.h.s. of \eqn{gravitylimitbcd},
which go as $s/t$,
contribute to $\cM^\Three$ to leading order in the Regge limit.
\para

Since the numerator factors for diagrams $a$, $b$, $c$, and $d$
are independent of loop momenta for both gauge theory and gravity, 
the corresponding integrals are 
simply proportional to the same integrals without any numerator factors
\begin{align}
I_{12jk}^\xone &= s^2 I_{12jk}^\xzero\,,
&
I_{13jk}^\xone &= u^2 I_{13jk}^\xzero 
~\longrightarrow ~s^2 I_{13jk}^\xzero \,,
&  \hbox{for}~ x &= a, b, c, d ,
\label{onezeroabcd}
\\[3mm]
I_{12jk}^\xtwo &= s^4 I_{12jk}^\xzero\,, &
I_{13jk}^\xtwo &= u^4 I_{13jk}^\xzero 
~\longrightarrow~ s^4 I_{13jk}^\xzero \,,
& \hbox{for}~ x &= a, b, c, d
\label{twozeroabcd}
\end{align}

which will be used below.
\para

\begin{figure}[t]
\begin{center}
\includegraphics[width=5.0cm]{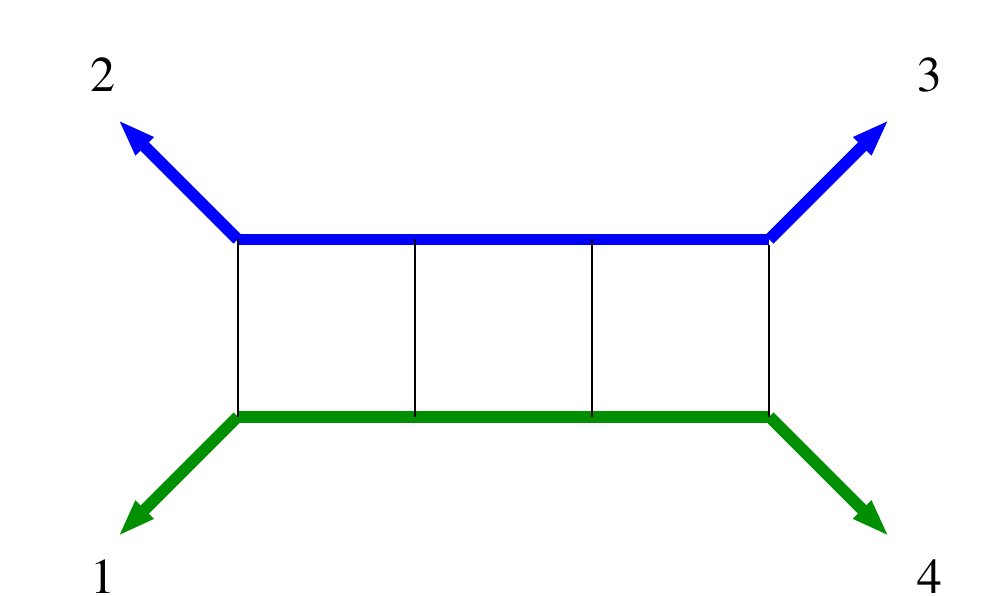}
\caption{Momentum routing for $I^{(a)}_{1234}$ }
\label{fig:a}
\end{center}
\end{figure}

\begin{figure}[b]
\begin{center}
\includegraphics[width=5.0cm]{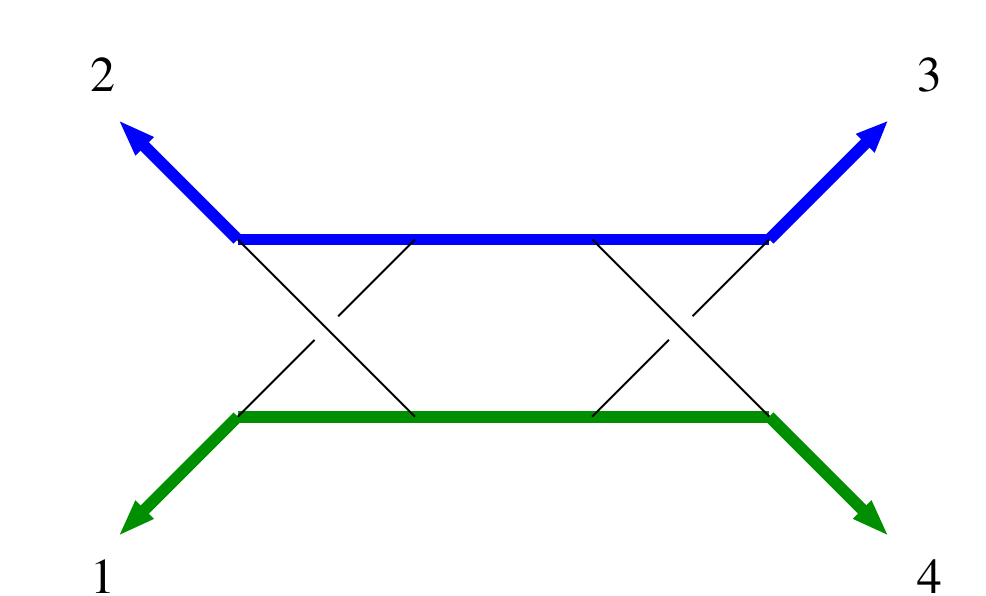}
\includegraphics[width=5.0cm]{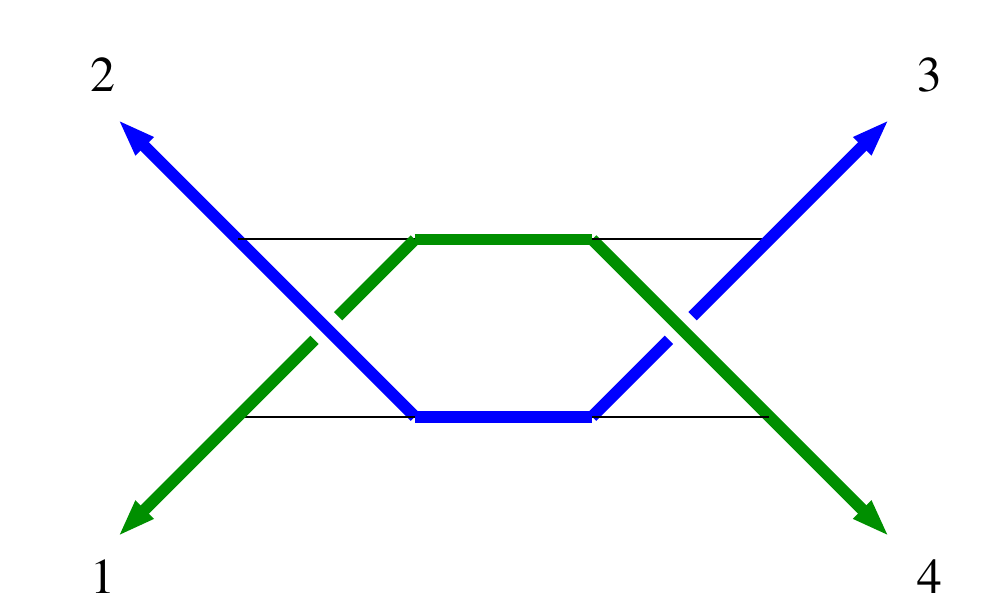}
\caption{Two possible momentum routings for $I^{(d)}_{1234}$ }
\label{fig:d}
\end{center}
\end{figure}

Now we turn to the more difficult integrals,
whose numerator factors
are dependent on the loop momenta.
To determine the Regge limits of these integrals,
we consider the routing of hard momenta 
through each of the integrals.  
Consider diagram $a$ shown in fig.~\ref{fig:a},
where we have adopted the convention that all 
external momenta are outgoing.
In the Regge limit 
$p_4  \to - p_1$ 
and 
$p_3 \to -p_2$,
and the hard momentum associated with each of these external legs
flows through the thick (green and blue) lines of the
diagram.
The thin (black) lines of the diagram carry 
the much softer exchanged momenta,
which we generically denote by $q$
in this paper. 
 In the Regge limit, $q^2$ is of order $t$.
Sometimes more than one routing of hard momenta
through the diagram is possible, 
as shown in fig.~\ref{fig:d} for the ``double-cross'' diagram $d$ .
This fact will play a key role in the discussion of ladder diagrams
in sec.~\ref{sec:ladders}.
\para

\begin{figure}[b]
\begin{center}
\includegraphics[width=4.0cm]{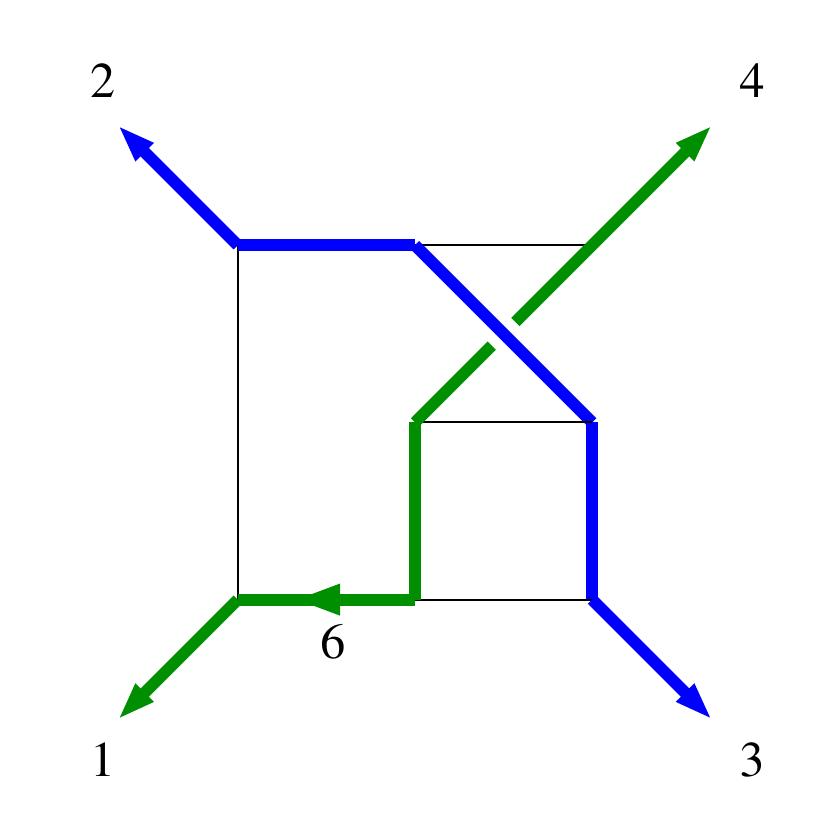}
\includegraphics[width=4.0cm]{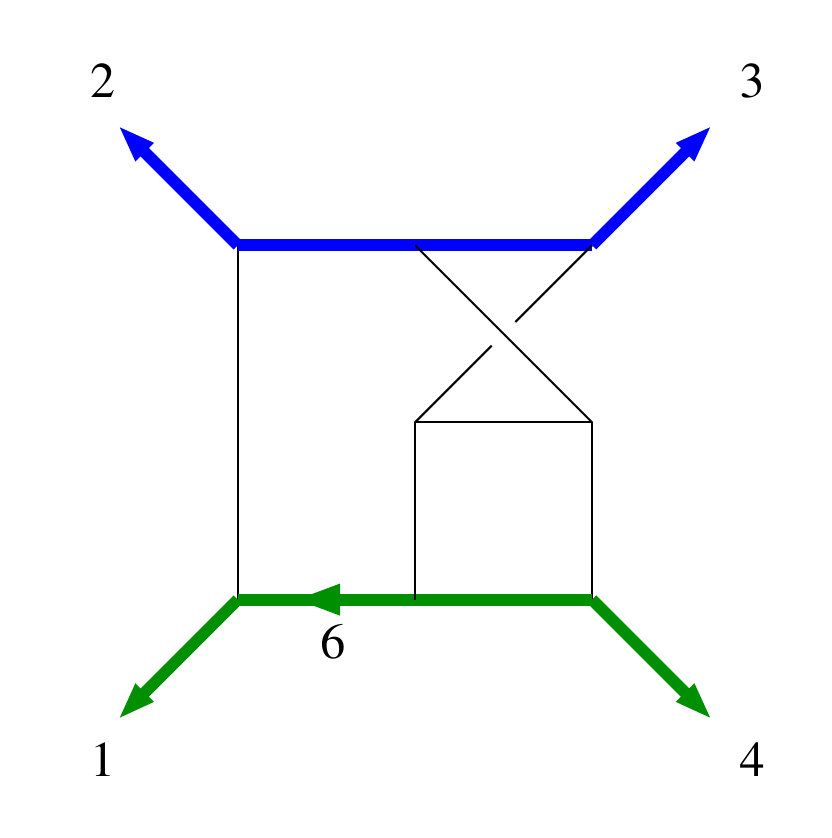}
\includegraphics[width=4.0cm]{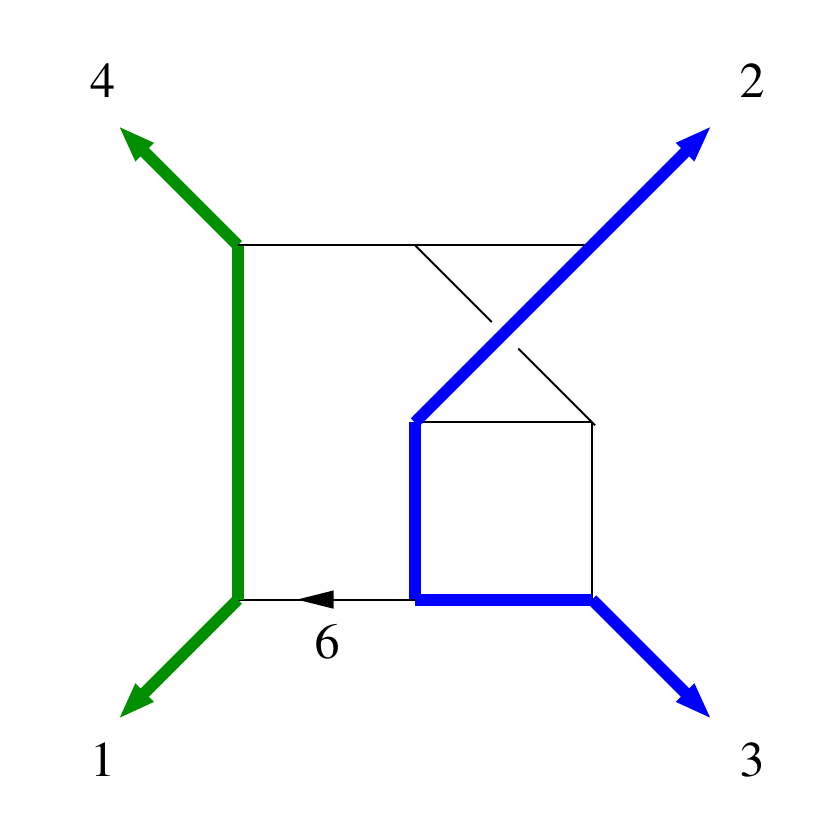}
\caption{Possible routings for
$I^{(f)}_{1243}$,
$I^{(f)}_{1234}$, and 
$I^{(f)}_{1423}$ respectively }
\label{fig:f}
\end{center}
\end{figure}

Now consider the routing of hard momenta through 
the gauge-theory integral 
$I_{1243}^{(f,1)} $,
shown in the first diagram of fig.~\ref{fig:f}. 
The numerator factor for this integral  is 
$s_{12} s_{36}$, where $s_{36} = (p_3 + l_6)^2$.
The momentum $l_6$ flowing through the thick (green) line 
in the direction indicated by the arrow
is approximately equal to $p_1$ 
(that is, $l_6 = p_1 + q$,
where $q$ is the soft momentum flowing through the line
connecting legs 1 and 2).
Hence in the Regge limit the numerator factor 
$s_{12} s_{36} \to s_{12} s_{13} \to - s^2$,
which we can then pull out of the integral.
The same reasoning obtains for $I_{1342}^{(f,1)} $,
and also for $I_{1243}^{(g,1)} $ and $I_{1342}^{(g,1)} $, 
so that 
\begin{align}
I_{1243}^{(f,1)} &\regge  - s^2 I_{1243}^{(f,0)}, 
& I_{1342}^{(f,1)} &\regge  - s^2 I_{1342}^{(f,0)},  \nn\\[2mm]
I_{1243}^{(g,1)} &\regge  - s^2 I_{1243}^{(g,0)}, 
&I_{1342}^{(g,1)} &\regge  - s^2 I_{1342}^{(g,0)}. 
\label{onezerofg}
\end{align}

Now consider the corresponding gravity integral 
$I_{1243}^{(f,2)} $.
The numerator factor for this integral is 
$(s_{12} s_{63})^2$,
which by the reasoning above goes to $s^4$ in the
Regge limit.
Using similar reasoning with the other integrals we obtain
\begin{align}
I_{1243}^{(f,2)} &\regge   s^4 I_{1243}^{(f,0)}, 
& I_{1342}^{(f,2)} &\regge   s^4 I_{1342}^{(f,0)}, \nn\\[2mm]
I_{1243}^{(g,2)} &\regge   s^4 I_{1243}^{(g,0)}, 
&I_{1342}^{(g,2)} &\regge   s^4 I_{1342}^{(g,0)} .
\label{twozerofg}
\end{align}

For the gravity integral $I_{1234}^{(f,2)} $,
one possible routing of hard momentum is shown in 
the second diagram of fig.~\ref{fig:f}.
The hard momentum continues to flow through 6, 
so $l_6 \to p_1$,
but the numerator factor in this case is 
$(s_{12} s_{46})^2 \to (s_{12} s_{14})^2 = s^2 t^2$, 
so this integral is suppressed relative to $I_{1243}^{(f,2)} $,
and similar reasoning holds 
$I_{1324}^{(f,2)} $,
and also for $I_{1234}^{(g,2)} $ and $I_{1324}^{(g,2)} $.
\para

For the gravity integral $I_{1423}^{(f,2)}$, 
one possible routing of hard momenta 
is shown in the third diagram of fig.~\ref{fig:f},
so that $l_6$ is soft.
Consequently, 
the numerator factor
$(s_{14} s_{36})^2  \to  t^4$,
suppressing this integral even further in the Regge limit.
Similar reasoning holds for 
$I_{1432}^{(f,2)} $,
and also for $I_{1423}^{(g,2)} $ and $I_{1432}^{(g,2)} $.
Thus, the  only permutations of the $f$ and $g$ integrals 
that contribute to $\cM^\Three$ in the Regge limit
are those shown in \eqn{twozerofg}.
 \para

Next, we consider diagram $e$.
The gravity integral $I_{1ijk}^{(e,2)}$
has numerator factor $(s_{1i} s_{k6})^2$,
compared to the gauge-theory numerator factor $s_{1i} s_{k6}$.
If $i\neq 4$, the routing of hard momentum goes through $l_6$,
so $l_6 \to p_1$  in which case
$s_{1i} s_{k6} \to s_{1i} s_{1k}$.
Thus in the Regge limit
\be
I^{(e,2)}_{1ijk} \regge 
s_{1i} s_{1 k} 
I^{(e,1)}_{1ijk} \,,
\qquad \hbox{for } i \neq 4 \,.
\ee

The kinematic prefactor for the gauge-theory integral 
$I^{(e,1)}_{1ijk} $
is given by \eqn{diagramae}, so we 
have 
\begin{align}
I^{(e,2)}_{1ijk} \regge 
{s_{1i} s_{1 k} \over 
s_{1i} s_{1 k} } \quad = \quad  1 \quad \ll \quad {s \over t} \,,
\qquad \hbox{for } i \neq 4
\end{align}

which is much less than the contribution of the other gravity diagrams.
The Regge limit of $I^{(e,1)}_{1ijk} $ 
with $i=4$ is even more highly suppressed (since $l_6$ is soft),
therefore diagram $e$
does not contribute to $\cM^\Three$ at all in the Regge limit.
\para

\begin{figure}[b]
\begin{center}
\includegraphics[width=4.0cm]{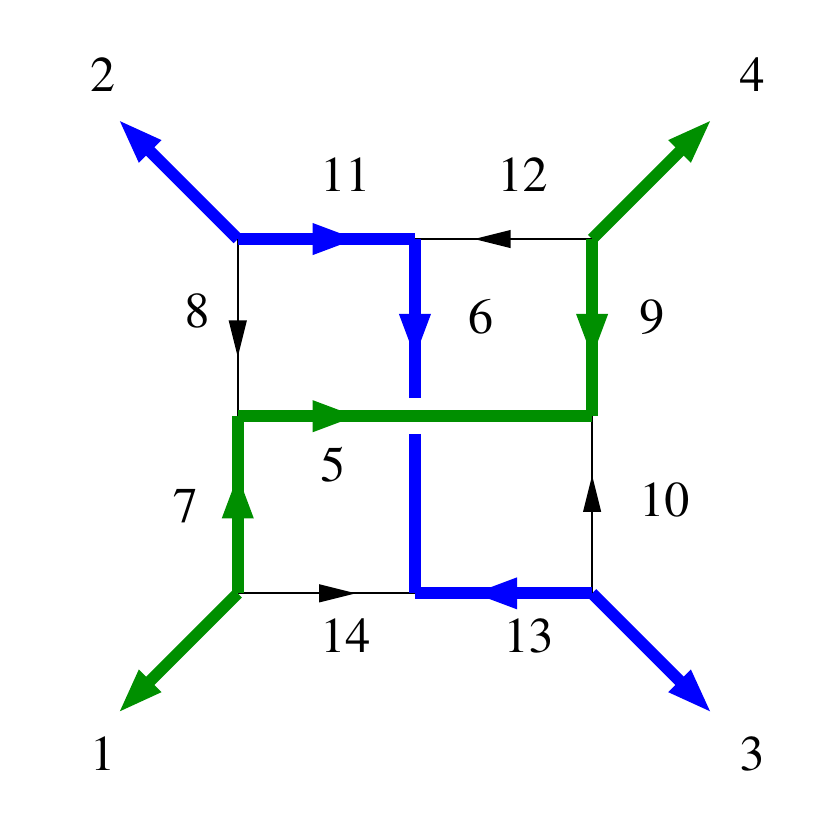}
\includegraphics[width=4.0cm]{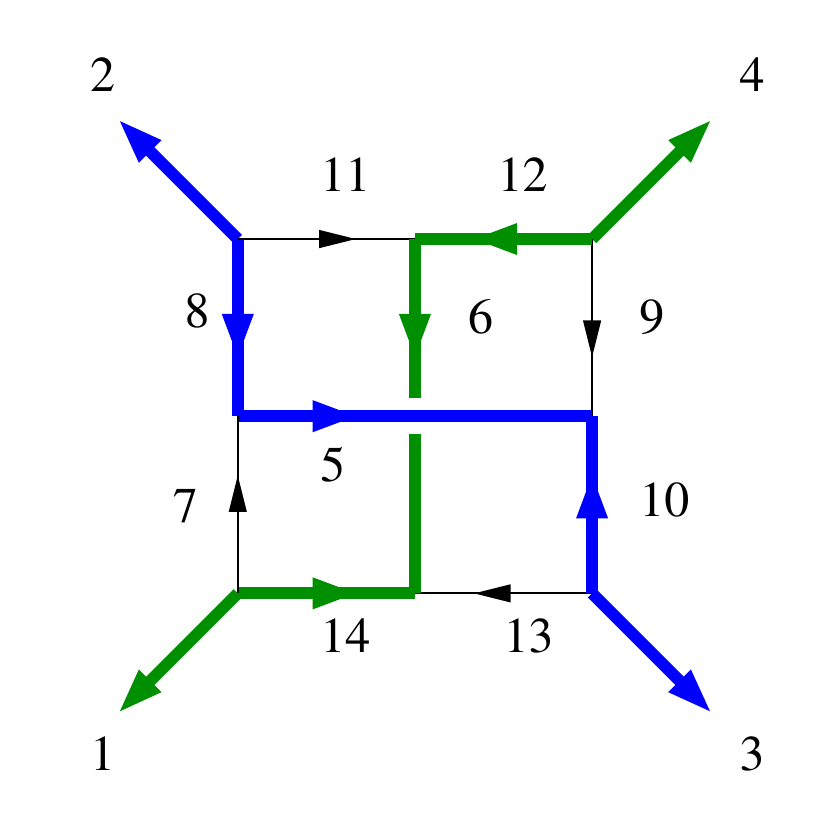}
\caption{Possible momentum routings for $I^{(h)}_{1243}$}
\label{fig:h}
\end{center}
\end{figure}

Finally we turn to the formidable integral $I_{1243}^{(h,1)} $,  
through which there are two possible routings of the hard momentum,
as shown in fig.~\ref{fig:h}. 
For the first routing, one has 
$l_5 \to p_4$ and $l_6 \to p_3$.
The numerator factor for this gauge-theory integral 
(cf. table 1)
simplifies in the Regge limit to 
\begin{align}
s_{12} (\tau_{26}+ \tau_{46}) 
+ s_{13} (\tau_{15}+ \tau_{25}) 
+ s_{12} s_{13}
&~\to~
s_{12} (s_{23}+ s_{34}) 
+ s_{13} (s_{14}+ s_{24}) 
+ s_{12} s_{13}
~\to~  s^2  \,. ~
\end{align}

In the second routing shown in fig.~\ref{fig:h},
one has $l_5 \to p_3$ and $l_6 \to p_1$,
and the numerator factor again approaches $s^2$
in the Regge limit.
Similar reasoning holds for $I_{1342}^{(h,1)}$.
Consequently, we see that the  
gauge-theory integrals corresponding to diagram $h$ go as 
\begin{align}
I_{1243}^{(h,1)} &\regge   s^2 I_{1243}^{(h,0)}, 
& I_{1342}^{(h,1)} &\regge   s^2 I_{1342}^{(h,0)} \,.
\label{onezeroh}
\end{align}

For the gravity integrals corresponding to diagram $h$,
the numerator factor (given in table 2) is more complicated.
In the Regge limit, however, most of the terms are subleading.
Any term $l_j^2$ goes either as $q^2$ 
(if the momentum flowing through leg is soft, $l_j \sim q$) 
or as $q \cdot p_i$
(if the momentum flowing through leg is hard, 
$l_j \sim p_i +q$, where $p_i$ is the momentum 
of one of the external legs),
both of which are $\ll s$.
Thus the gravity numerator factor for
$I_{1243}^{(h,2)} $
simplifies to 
\be
(s_{12} s_{89} + s_{13} s_{11,14} - s_{12} s_{13})^2 
~\to~  s^4
\ee

because $s_{89}$ and $s_{11,14}$ are both $\sim q \cdot p_i$.
Consequently we may write 
\begin{align}
I_{1243}^{(h,2)} &\regge   s^4 I_{1243}^{(h,0)}, 
& I_{1342}^{(h,2)} &\regge   s^4 I_{1342}^{(h,0)} \,.
\label{twozeroh}
\end{align}

Straightforward consideration of all 
other permutations of the external legs
for this diagram shows that 
the corresponding gravity integrals
are subleading in the Regge limit.
The same is true for the gravity integrals
corresponding to all permutations 
of external legs for diagram $i$.
(Note that diagram $i$ does not contribute
at all to the gauge theory expression (\ref{threelooprhs}).)
\para

We are now in a position to combine all of the results obtained above
to evaluate both 
\eqns{threelooprhs}{threeloopgrav} 
in the Regge limit.
Taking into account \eqn{limita},
and using eqs.~(\ref{onezeroabcd}),
(\ref{onezerofg}), and (\ref{onezeroh}),
we find that the linear combination of 
gauge-theory amplitudes (\ref{threelooprhs}) 
evaluates in the Regge limit to 
\begin{align} 
&\hskip-30mm
{
  4 \left(  A_1^\Three + A_3^\Three \right) 
- \left( A_4^\Three +A_6^\Three + A_7^\Three +A_9^\Three \right)
 \over
g_D^6  s t A_1^\Zero  }
\nn\\
\regge 
12 s^2 \Big[  
& 
I_{1234}^{(a,0)} 
+ I_{1324}^{(a,0)}
+ 2 I_{1234}^{(b,0)} 
+2 I_{1342}^{(b,0)}
+ \fr{1}{2}  \left(  
2 I_{1234}^{(c,0)} 
+2 I_{1342}^{(c,0)}
\right)
+ \fr{1}{4}  \left(  
 2 I_{1234}^{(d,0)} 
+2 I_{1342}^{(d,0)}
\right)
\nn\\ &
+2  \left( 
 I_{1243}^{(f,0)}
+ I_{1342}^{(f,0)}
\right)
+4 \left(
  I_{1243}^{(g,0)}
+ I_{1342}^{(g,0)}
\right)
+\fr{1}{2}  \left(
  I_{1243}^{(h,0)}
+ I_{1342}^{(h,0)}
\right)
\Big].
\label{finalgaugeregge}
\end{align}

Similarly, we can evaluate the gravity amplitude (\ref{threeloopgrav}) 
by omitting all the integrals suppressed in the Regge limit,
and using eqs.~(\ref{twozeroabcd}),
(\ref{twozerofg}), and (\ref{twozeroh})
to obtain 
\begin{align}
&\hskip-27mm
 {1 \over ( \kappa_D/2)^6    ~s t u }
 {\cM^\Three \over \cM^\Zero }
\nn\\
\regge
s^4 \Big[
&
 I_{1234}^{(a,0)} 
+I_{1324}^{(a,0)}
+2I_{1234}^{(b,0)}
+2I_{1342}^{(b,0)}
+\fr{1}{2}  \left( 
 2 I_{1234}^{(c,0)}
+2I_{1342}^{(c,0)}
\right)
+\fr{1}{4}  \left(
 2I_{1234}^{(d,0)}
+2 I_{1342}^{(d,0)}
\right)
\nn\\
&
+2 \left(
  I_{1243}^{(f,0)}
+ I_{1342}^{(f,0)}
\right)
+4 \left(
  I_{1243}^{(g,0)}
+ I_{1342}^{(g,0)}
\right)
+\fr{1}{2}  \left(
  I_{1243}^{(h,0)}
+ I_{1342}^{(h,0)}
\right)
\Big] .
\label{threeloopladders} 
\end{align}

Comparing \eqns{finalgaugeregge}{threeloopladders},
we immediately see that the conjectured three-loop SYM/supergravity 
relation (\ref{threelooprelationbis})
is confirmed in the Regge limit. 

\subsection{An exact three-loop relation?}

It might be asked whether the three-loop relation (\ref{threelooprelationbis})
proved in the previous subsection is a specialization to the Regge 
limit of some exact relation, as was the case at one and two loops.
Because the three-loop kinematic numerators for both 
supergravity and SYM theory are dependent on the loop momenta
for diagrams (e) through (i) of fig.~\ref{IntegralsThreeLoopFigure}
(see tables 1 and 2),
 the double-copy procedure,
which relates the integrands of these amplitudes, 
does not give a straightforward relation 
between the integrated amplitudes
(unlike the one- and two-loop cases).
\para

Interestingly, if one were to include {\it only} diagrams (a) through (d) 
of fig.~\ref{IntegralsThreeLoopFigure}
(namely, those with numerator factors that are independent of loop momenta)
in the expressions for the three-loop amplitudes 
(\ref{threeloopyangmills}) and (\ref{threeloopgrav})
then one can derive a unique exact (i.e. not only in the Regge limit) relation 
between them, namely\footnote{Note that 
$A_\lam^\Three$ with $\lam=5, 10, 11,$ and 12 do not appear in this
relation because they have been eliminated using the three-loop group
theory relations;   see eq.~(2.13) of ref.~\cite{Naculich:2020clm}.}
\begin{align}
 {1 \over ( \kappa_D/2)^6    ~s t u }
 {\cM^\Three \over \cM^\Zero }
\mathrel{\mathop{=}\limits^{?!} } 
\Big[
&  2 (s^2 + u^2 - 2t^2) 
  \left(  A_1^\Three + A_2^\Three + A_3^\Three \right) 
- (u^2-t^2)  A_4^\Three 
- (s^2-t^2) A_6^\Three 
\nn\\
& 
- (u^2 + 2t^2)  A_7^\Three 
- 3t^2  A_8^\Three 
- (s^2 + 2t^2)  A_9^\Three 
\Big] \Big/ 
\left( 12 g_D^6  s t A_1^\Zero   \right) \,.
\label{exactthreelooprelation}
\end{align}

Moreover this relation indeed reduces to 
\eqn{threelooprelationbis} in the Regge limit.
\para

One might then ask whether the relation (\ref{exactthreelooprelation})
could be valid for the full amplitudes, 
despite the fact that it was obtained using only some
of the contributing diagrams.
Henn and Mistlberger have computed the Laurent expansions 
(through $\cO(\de^0)$)
of the three-loop 
$\cN=8$ supergravity amplitude,
which starts at $\cO(1/\de^3)$, 
in ref.~\cite{Henn:2019rgj}
and the color-ordered amplitudes of the 
three-loop $\cN=4$ SYM amplitude,
which start at $\cO(1/\de^6)$,
in ref.~\cite{Henn:2016jdu}.
Using their results, one finds,
somewhat remarkably, 
that the relation (\ref{exactthreelooprelation})
is satisfied through $\cO(1/\de^2)$,
with the difference between the two sides given by the rather simple expression
\be
\left( \hbox{l.h.s.} - \hbox{r.h.s} \right)_{\eqn{exactthreelooprelation}} 
= 
{1 \over (8 \pi^2)^3} { \zeta_5 + 2 \zeta_2 \zeta_3 \over \de}
+ \cO(\de^0) \,.
\label{breakdown}
\ee

Thus, 
while the relation (\ref{exactthreelooprelation})
is not generally valid (except in the Regge limit),
it holds better than one might have anticipated.
Note that the leading kinematical dependence on both sides of 
\eqn{exactthreelooprelation} goes as $\sim s/t$ in the Regge limit
so that the discrepancy (\ref{breakdown}) 
is subleading in an expansion in $t/s$.
It would be interesting to understand the reason for this discrepancy.

\section{All-loop order Regge limit of gravity amplitudes} 
\setcounter{equation}{0}
\label{sec:ladders}

In this section, 
we show that $\cN=8$ supergravity four-point amplitudes
at one, two, and three loops reduce in the Regge limit 
to a (modified) sum of ladder and crossed-ladder scalar diagrams,
and explain how this is related to the eikonal representation
of gravity amplitudes. 

\subsection{Ladder and crossed-ladder diagrams}

\begin{figure}[b]
\begin{center}
\includegraphics[width=6.0cm]{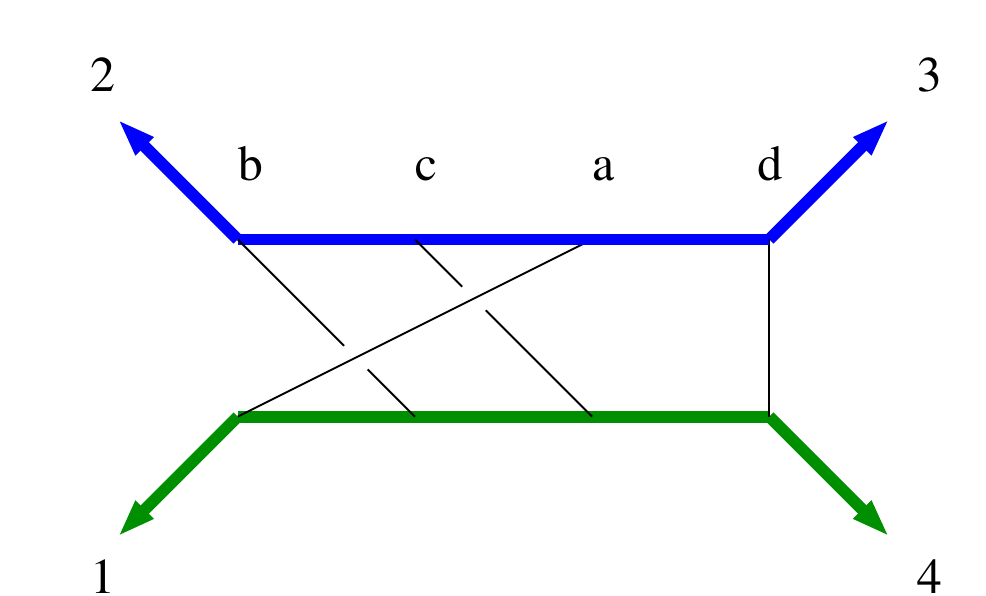}
\caption{Crossed-ladder diagram corresponding to the integral $I^\lad_{[abcd]}$.}
\label{fig:xlad}
\end{center}
\end{figure}

The expressions for the Regge limits of the 
$\ell$-loop $\cN=8$ supergravity amplitudes 
(and therefore the equivalent linear combinations of
color-ordered $\ell$-loop $\cN=4$ SYM amplitudes)
obtained in the previous three sections
take a very specific form:
they are (almost) the sum over all 
ladder and crossed-ladder scalar diagrams
at that loop order.
To show this,
we introduce a special notation for (crossed-)ladder diagrams
(see fig.~\ref{fig:xlad}).
The $\ell$-loop crossed-ladder integral $I^\lad_{[abc\ldots]}$ 
is defined with precisely the same prefactors 
(and no numerator factors) as $I^\xzero$ (cf. \eqn{Ixzero}),
with the subscript in brackets describing how the rungs are connected
between the rails of the ladder:
the first vertex in the thick (green) line running from
$1$ to $4$ is attached to the $a$th vertex in the thick (blue) line 
running from 2 to 3, the second vertex to the $b$th vertex, etc.
The specific diagram shown  in fig.~\ref{fig:xlad} 
corresponds to $I^\lad_{[3124]}$.
\para

The one-loop supergravity amplitude in the Regge limit, 
\eqn{oneloopladders},
is easily seen to be the 
sum of a ladder and a crossed ladder diagram
\be
\cM^\One 
\regge
\left( \kappa_D\over 2 \right)^2 
\,
stu
\cM^\Zero 
\left[ 
I_{[12]}^\lad  + I_{[21]}^\lad  \right] \,.
\label{oneladder}
\ee

The two-loop supergravity amplitude in the Regge limit, 
\eqn{twoloopladders},
which is equivalent by 
\eqns{klein}{twoloopnonplanarsymmetry} to 
\be
\cM^\Two 
\regge
-  \left( \kappa_D \over 2\right)^4   
 ~ s t u \cM^\Zero 
s^2  \Big[ 
 I_{1234}^\Pzero
+I_{1324}^\Pzero
+ I_{1234}^\NPzero
+ I_{4321}^\NPzero
+ I_{1324}^\NPzero
+ I_{4231}^\NPzero
\Big] 
\ee

is written in the ladder notation as 
\be
\cM^\Two 
\regge
-  \left( \kappa_D \over 2\right)^4   
 ~ s t u \cM^\Zero 
s^2  \Big[ 
 I_{[123]}^\lad
+I_{[321]}^\lad
+I_{[132]}^\lad
+I_{[213]}^\lad
+I_{[312]}^\lad
+I_{[231]}^\lad 
\Big]
\label{twoladder}
\ee
in which all six permutations of the three rungs are present.
(This was observed in ref.~\cite{Giddings:2010pp}.)
\para

Finally, we recall from \eqn{threeloopladders}
the three-loop supergravity amplitude in the Regge limit
\begin{align}
\cM^\Three
&\regge
\left( \kappa_D \over 2\right)^6   
 ~ s t u \cM^\Zero 
s^4  \Big[ 
 I_{1234}^{(a,0)} 
+I_{1324}^{(a,0)}
+2I_{1234}^{(b,0)}
+2I_{1342}^{(b,0)}
\nn\\
&
+\fr{1}{2}  \left( 
 2 I_{1234}^{(c,0)}
+2I_{1342}^{(c,0)}
\right)
+\fr{1}{4}  \left(
 2I_{1234}^{(d,0)}
+2 I_{1342}^{(d,0)}
\right)
+2 \left(
  I_{1243}^{(f,0)}
+ I_{1342}^{(f,0)}
\right)
\nn\\
&
+4 \left(
  I_{1243}^{(g,0)}
+ I_{1342}^{(g,0)}
\right)
+\fr{1}{2}  \left(
  I_{1243}^{(h,0)}
+ I_{1342}^{(h,0)}
\right)
\Big] .
\label{finalgravityregge}
\end{align}

Using the invariance of 
the integrals under the Klein four-group (\ref{klein}), 
this may be recast as 
\begin{align}
\cM^\Three
&
\regge
\left( \kappa_D \over 2\right)^6   
 ~ s t u \cM^\Zero 
s^4  \Big[ 
 I_{1234}^{(a,0)} 
+I_{1324}^{(a,0)}
+ I^{(b,0)}_{1234} 
+ I^{(b,0)}_{4321} 
+ I^{(b,0)}_{1324} 
+ I^{(b,0)}_{4231} 
\nn\\ 
&
+ I_{1234}^{(c,0)}
+I_{1324}^{(c,0)}
+\fr{1}{2}  \left(
 I_{1234}^{(d,0)}
+I_{1324}^{(d,0)}
\right)
+ I_{1243}^{(f,0)}
+ I_{4312}^{(f,0)} 
+ I_{1342}^{(f,0)}
+ I_{4213}^{(f,0)} 
\label{threeloopladdersemifinal}
\\ 
&
+ I^{(g,0)}_{1243} + I^{(g,0)}_{2134} + I^{(g,0)}_{3421} + I^{(g,0)}_{4312} 
+ I^{(g,0)}_{1342} +  I^{(g,0)}_{2431} + I^{(g,0)}_{3124} + I^{(g,0)}_{4213}
+\fr{1}{2}  \left(
  I_{1243}^{(h,0)}
+ I_{1342}^{(h,0)}
\right)
\Big] \,.
\nn
\end{align}

We recognize all of these integrals as corresponding 
to ladders and crossed ladders
\begin{align}
\cM^\Three
&
\regge
\left( \kappa_D \over 2\right)^6   
 ~ s t u \cM^\Zero 
s^4  \Big[ 
 I_{[1234]}^\lad
+I_{[4321]}^\lad
+I_{[1243]}^\lad
+I_{[2134]}^\lad
+I_{[4312]}^\lad
+I_{[3421]}^\lad
\nn\\ &
+I_{[1324]}^\lad
+I_{[4231]}^\lad
+\fr{1}{2}  \left(
 I_{[2143]}^\lad
+I_{[3412]}^\lad
\right)
+I_{[1432]}^\lad
+I_{[3214]}^\lad
+I_{[4123]}^\lad
+I_{[2341]}^\lad
\label{threeladder}
\\ &
+I_{[1423]}^\lad
+I_{[1342]}^\lad
+I_{[3124]}^\lad
+I_{[2314]}^\lad
+I_{[4132]}^\lad
+I_{[2431]}^\lad
+I_{[4213]}^\lad
+I_{[3241]}^\lad
+
I_{[3142]}^\lad
+I_{[2413]}^\lad
\Big] \,.
\nn
\end{align}

One slightly subtle point deserves to be noted
in going from  \eqn{threeloopladdersemifinal}
to \eqn{threeladder}.
Each of the integrals
$ I_{1243}^{(h,0)} $ 
and $ I_{1342}^{(h,0)}$
has two possible momentum routings, 
as shown in fig.~\ref{fig:h}.
Each routing contributes to separate crossed-ladder integrals,
$ I_{[3142]}^\lad$ and $I_{[2413]}^\lad$,
which accounts for the disappearance of the 
factor of $1/2$ multiplying the last pair of terms in 
\eqn{threeloopladdersemifinal}.
\para

Examination of \eqn{threeladder}
shows that it contains all 24 permutations of 
the rungs of the crossed ladders,
{\it except} that two of them include a factor of 1/2. 
This is a blessing in disguise,
as we will now see.
\para

\subsection{Eikonal representation} 

An alternative approach to evaluating the Regge limit 
of supergravity uses the eikonal 
approximation \cite{Cheng:1969eh,Abarbanel:1969ek,Levy:1969cr}
to write the gravitational amplitude (to all loop orders) 
in impact-parameter space
\cite{tHooft:1987vrq,Amati:1987wq,Muzinich:1987in,Amati:1987uf,Kabat:1992tb,
Giddings:2010pp,Melville:2013qca,Akhoury:2013yua,Luna:2015paa,
DiVecchia:2019myk,DiVecchia:2019kta}
\be
\cM \sim 
\int d^{D-2}\vec{x}_\perp
e^{-i\vec{q}_\perp\cdot \vec{x}_\perp} \left(e^{i\chi(\vec{x}_\perp)}-1
\right)
\label{Meik}
\ee

where $\vec{x}_\perp$ is a $(D-2)$-dimensional vector
transverse to the incoming particle direction,
and $\vec{q}_\perp$ is the $(D-2)$-dimensional
momentum transfer that is Fourier-conjugate to $\vec{x}_\perp$ 
(where $t\simeq -|\vec{q}_\perp|^2$ in the leading Regge limit).
The quantity $i\chi(\vec{x}_\perp)$ is known as the eikonal phase, 
and is given in $D= 4- 2\epsilon$ dimensions by
\be
i\chi(\vec{x}_\perp)
= {-iG_D s \over \de} 
\Gamma(1-\de) (\pi\vec{x}_\perp^2)^\epsilon \,.
\label{chidef}
\ee

By expanding the exponential in \eqn{Meik}
in a Taylor series in $G_D$
and Fourier transforming one 
obtains \cite{DiVecchia:2019myk}
\be
\cM^\Ell \regge \cM^\Zero 
{1 \over \ell!}  
\left[ 
{ \Gamma^2(1-\eps) \Gamma(1+\eps) \over \Gamma(1-2\eps)}
\left( 4 \pi  \over -t \right)^{\de}  
\left( - i G_D s \over \de \right)
\right]^\ell
G^\Ell(\de) 
\label{Mell}
\ee

where
\be
G^\Ell (\de)
=
\frac{\Gamma^\ell (1-2 \de) \Gamma (1+\ell \de)}
{\Gamma^{\ell-1} (1 -\de) \Gamma^\ell (1+ \de) \Gamma (1- (\ell+1) \de)} \,.
\ee

As explained in ref.~\cite{DiVecchia:2019myk},
this result is consistent with the Regge limits of the 
Laurent expansions of the one-, two-, 
and three-loop $\cN=8$ supergravity amplitudes
obtained in ref.~\cite{Henn:2019rgj}.
(A proposed extension \cite{DiVecchia:2019kta} 
of the eikonal representation (\ref{Meik}) 
to include subleading-level, i.e. $\cO(-t/s)$, 
corrections also agrees with the results of ref.~\cite{Henn:2019rgj},
up to a small discrepancy at the three-loop level
that has still not been fully accounted for.) 
\para

What do \eqns{Meik}{Mell} have to do with 
the expressions for $\cN=8$ supergravity amplitudes
in terms of scalar integrals (without numerator factors)
obtained in 
eqs.~(\ref{oneladder}), (\ref{twoladder}), and (\ref{threeladder}) 
above?
Starting in the 1960's,
efforts were made to derive the eikonal representation (\ref{Meik})
from the sum of ladder and crossed-ladder scalar 
diagrams \cite{Levy:1969cr,Tiktopoulos:1970nr}.
(See appendix C of ref.~\cite{Cheng:1987ga}
for a detailed accounting.)
While successful at one and two loops,
this project was discovered to fail at three loops and above
\cite{Tiktopoulos:1971hi,Cheng:1987ga,Kabat:1992pz}. 
At three loops, the obstacle is the fact 
that there are two possible routings of hard momentum 
through the ``double-cross'' ladder diagram $d$
(as shown in fig.~\ref{fig:d})
so that the contribution from this integral in the Regge limit
is twice what it needs to be to give the eikonal result (\ref{Mell}).
\para

We found above, however, that the double-cross ladder diagrams
that appear in the Regge limit of the three-loop 
$\cN=8$ supergravity amplitude (\ref{threeladder}) 
obtained
using generalized unitarity in refs.~\cite{Bern:2007hh,Bern:2008pv}
come equipped with
a factor of 1/2, 
which precisely corrects for this overcounting.
Thus, in the Regge limit,
the (modified) sum of ladders and crossed ladders given in \eqn{threeladder}
yields the (correct) eikonal three-loop result (\ref{Mell}).

\section{Conclusions}
\setcounter{equation}{0}
\label{sec:concl}

In a previous paper, one of the authors analyzed the 
structure of the Regge limit of the (nonplanar) 
$\cN=4$ SYM four-point amplitude \cite{Naculich:2020clm}, 
and based on those results,
conjectured an all-loop-orders relation 
between the Regge limits of 
the four-point amplitudes of
$\cN=4$ SYM theory and $\cN=8$ supergravity,
viz. \eqns{evenloop}{oddloop}.
The one- and two-loop Regge limit relations,
\eqns{onelooprelation}{twolooprelation},
are consequences of known exact relations,
\eqns{exactonelooprelation}{exacttwolooprelation},
between
$\cN=4$ SYM and $\cN=8$ supergravity amplitudes.
\para

In this paper, we established the conjectured 
(Regge limit) relation at the three-loop level,
viz. \eqn{threelooprelation}.
We showed that the Regge limit of exact expressions 
for the amplitudes, obtained using generalized unitarity,
simplifies in both cases to the same (modified) sum over 
three-loop ladder and crossed-ladder scalar diagrams,
thus proving the conjectured relation.
The sum is modified in the sense that two of the 
crossed-ladder diagrams are multiplied by a factor of one-half
relative to the remaining diagrams.
\para

We also presented an exact three-loop relation 
(\ref{exactthreelooprelation})
that would be valid if only a certain subset of the scalar diagrams
were included in the evaluation of the three-loop amplitudes,
and which reduces to \eqn{threelooprelation} in the Regge limit.   
We tested this putative relation against the Laurent expansions
of the full three-loop amplitudes,
and found rather remarkably that it holds 
at $\cO(1/\de^3)$ and $\cO(1/\de^2)$,
and only breaks down at $\cO(1/\de)$.
\para

The supergravity four-point amplitude can alternatively 
be evaluated in the Regge limit 
using the eikonal approximation to give 
a representation in impact-parameter space (\ref{Meik}).
This may in turn be evaluated \cite{DiVecchia:2019myk}
to give the expression (\ref{Mell}),
and shown to agree with the known 
Regge limit of the $\cN=8$ supergravity amplitude
through three loops.
\para

In this paper, 
we showed that the modification of the 
sum over crossed-ladder scalar diagrams
described above is precisely what is required
for the Regge limit of the sum
to agree with correct three-loop result,
whereas it is known that the unmodified sum
fails to do so \cite{Tiktopoulos:1971hi,Cheng:1987ga,Kabat:1992pz}.

\section*{Acknowledgments}
This material is based upon work supported by the
National Science Foundation
under Grant No.~PHY21-11943.

\vfil\break

\end{document}